\documentclass[twocolumn,letterpaper,pra,showpacs,superscriptaddress]{revtex4}
\newlength{\twocolumnwidth}

\usepackage{graphicx}
\usepackage{bm}
\usepackage{amsmath}
\usepackage{ifthen}
\usepackage{psfrag}
\begin{document}
\title{Commuting Heisenberg operators as the quantum response problem:\\
Time-normal averages in the truncated Wigner representation.}
\author{B.~Berg}
\affiliation{Institut f\"ur Quantenphysik, Universit\"at Ulm, D-89069, Ulm, Germany.}
\author{L.~I.~Plimak}
\affiliation{Institut f\"ur Quantenphysik, Universit\"at Ulm, D-89069, Ulm, Germany.}
\affiliation{ARC Centre of Excellence for Quantum-Atom Optics, School of Physical Sciences, University of Queensland, Brisbane, Qld 4072, Australia.}
\thanks{{\em E-mail\/}: Bettina.Berg@uni-ulm.de.}
\author{A.\ Polkovnikov}
\affiliation{Boston University, Department of Physics, Boston, MA, 02215, USA}
\author{M.~K.~Olsen}
\affiliation{Institut f\"ur Quantenphysik, Universit\"at Ulm, D-89069, Ulm, Germany.}
\affiliation{ARC Centre of Excellence for Quantum-Atom Optics, School of Physical Sciences, University of Queensland, Brisbane, Qld 4072, Australia.}
\author{M.~Fleischhauer}
\affiliation{Fachbereich Physik, Technische Universit\"at Kaiserslautern, D-67633 Kaiserslautern, Germany.}
\author{W.~P.~Schleich}
\affiliation{Institut f\"ur Quantenphysik, Universit\"at Ulm, D-89069, Ulm, Germany.}
\date{\today}
\begin{abstract}
The applicability of the so-called truncated Wigner approximation (--W) is extended to multitime averages of Heisenberg field operators. This task splits naturally in two. Firstly, what class of multitime averages the --W approximates, and, secondly, how to proceed if the average in question does not belong to this class. To answer the first question we develop an (in principle, exact) path-integral approach in phase-space based on the symmetric (Weyl) ordering of creation and annihilation operators. These techniques calculate a new class of averages which we call {\em time-symmetric\/}. The --W equations emerge as an approximation within this path-integral techniques. We then show that the answer to the second question is associated with response properties of the system. In fact, for two-time averages Kubo's renowned formula relating the linear response function to two-time commutators suffices. The --W is trivially generalised to the response properties of the system allowing one to calculate approximate time-normally ordered two-time correlation functions with surprising ease. The techniques we develop are demonstrated for the Bose-Hubbard model.

\end{abstract}
\maketitle

\section{Introduction}
One of the most fascinating areas of ultra-cold atomic physics is
the experimental investigation of the dynamical properties
of interacting many-body systems. The control of experimental parameters and ability to tailor systems is allowing many interesting effects to be observed which would have been almost impossible in the recent past.
Among important examples are the recently observed dynamical instabilities~\cite{Fallani} and inhibited transport~\cite{Fertig, mun} in one dimensional (1D) and three-dimensional (3D) optical lattice systems, as well as nonlinear self-trapping in 1D periodic potentials~\cite{Anker} and in Josephson junctions~\cite{Albiez}. Despite these important experimental advances, the theoretical description of dynamical properties, especially for strong
interactions, remains a major challenge, with progress having been incremental up to now. For bosons, except for rare cases where the dynamics are analytically tractable, e.g. by Bethe ansatz~\cite{Caux}, one method is to adapt the phase-space representations of quantum optics~\cite{Glauber,Sudarshan,plusP,GardinerQN}. Attempts have been made, for example, to apply the positive-P representation to the dynamics of trapped and colliding Bose-Einstein condensates~\cite{WBEC1,Joel,jjhletter,Deuar}, although these have been only partially successful due to numerical instabilities which mean that integration is restricted to either short times or small interaction strengths. We note here that efforts are being made to extend the usefulness of this method~\cite{gauge1,gauge2,gauge3}, with promising results for single-mode systems. For 1D systems numerical algorithms based on adaptive matrix-product decompositions of the state vector, such as the t-DMRG~\cite{DMRG} and the related TEBD algorithm~\cite{TEBD}, have recently been developed. They too are restricted however to short times or systems with slow entanglement growth.

For highly nonlinear underdamped systems such as trapped condensates,
it is often easier to use the truncated Wigner approximation (-W)~\cite{Wminus}, with the main advantage being that it is numerically stable~\cite{WBEC1,WBEC2,WBEC3,WBEC4} and simple to implement. The --W takes into account initial uncertainty between conjugate variables or initial quantum noise~\cite{ap_twa}, which is often necessary to trigger certain dynamical processes. This method was recently used to predict and explain several experiments in the context of interacting cold atom systems. As an example, using this method the damping of a dipolar motion of a 1D condensate in an optical lattice was first predicted in ref.~\cite{pw} and later results were qualitatively confirmed in experiment~\cite{Fertig}. This experiment was later simulated more accurately using a multi-band version of the --W~\cite{Ruostekoski}. In Refs.~\cite{Bistritser, Hutchinson} the --W was used to analyze splitting or merging between elongated condensates, closely mimicking the situation realized in recent experiments~\cite{joerg1}. In Ref.~\cite{psg}, the --W enabled the authors to explain the coherence dynamics after a sudden quench of tunneling from an insulator to a superfluid, giving good agreement with experimental results~\cite{ari}. For a fuller review of other recent developments, we direct the reader to Ref.~\cite{blakie}. 

While the truncated Wigner representation is becoming increasingly utilized, there are two drawbacks which prevent it from being the numerical technique of choice. The first is its approximate nature, arising from the truncation procedure, which may sometimes lead to demonstrably wrong results~\cite{Wigwrong1,Wigwrong2}. In principle, this drawback can be overcome. A way to fully map the quantum master equation onto Wigner representation stochastic difference equations has been developed, but does not result in a widely and easily applicable method~\cite{nossoEPL,ourpreprint,spontaneous}. An expansion allowing one to include the dropped contribution perturbatively via quantum jumps was suggested in Ref.~\cite{ap_twa}, but until now this has been restricted to the calculation of single-time averages.

The second drawback is that averages of the phase-space variables in the Wigner representation do not map directly onto expectation values of
{\em time-normally ordered\/} multitime operator products corresponding with experimental measurements. For two operators, we recover a symmetrised product, while for three and more operators, there emerges a new type of ordering of Heisenberg operators which we term {\em time-symmetric\/} \cite{ourpreprint}. Converting these to time-normal ordering requires the calculation of commutators of Heisenberg operators at {\em different times\/}. At first glance this drawback appears to be fundamental, but we will demonstrate that it can be circumvented with surprising ease, by making use of the fact that commutators of
Heisenberg operators at different times express the {\em response\/}
properties of a quantum system~\cite{ourpreprint,Kubo1,Kubo2,Corr,BWO,PlSt}.
Generally, commuting the Heisenberg operators requires solving the full
quantum stochastic nonlinear response problem, and, although possible in principle, this remains a formidable task \cite{PlSt,Unpb}. For two-time commutators, however, the process is simplified due to Kubo's linear response formulation~\cite{Kubo1,Kubo2}, which has previously been used to convert normally-ordered correlations into symmetrically ordered ones within the positive-P representation~\cite{Wigwrong2,QNDBWO}. In this paper we show how,
combined with the truncated Wigner representation as a computational tool, Kubo's linear response theory turns into a simple yet powerful approximate method of calculating two-time correlation functions of interacting bosonic fields with arbitrary operator ordering. We stress here that the Kubo relation is exact, with the approximate nature coming from the truncation process used to find the appropriate stochastic equations.

In more traditional phase-space techniques, the truncated-Wigner equations are developed by dropping the third-order derivatives in the generalised Fokker-Planck equation for the single-time Wigner distribution \cite{GardinerQN,Wminus}. The corresponding Langevin equations are either non-stochastic (without losses) or probabilistic (with losses) and are very easy to handle numerically. However simple and straightforward, this conventional way of deriving the truncated Wigner approach leaves it unclear if it can be applied to any {\em multitime\/} quantum averages. In our approach, the truncated-Wigner equations emerge as an approximation within rigorous phase-space path-integral techniques. By itself, the path integral expresses the averages of {\em time-symmetrically-ordered products of Heisenberg operators.\/}
The generalisation associated with extending the truncated-Wigner equations to multitime averages is thus highly nontrivial and requires a new concept: the time-symmetric ordering of the Heisenberg operators.
There does not seem to be a way of guessing this concept from within the conventional phase-space techniques. Just proving the equivalence between the time-symmetric and the conventional symmetric ordering of free-field operators is a nontrivial task~\cite{Unpb}. 

To make this paper more accessible we will begin with a theoretical summary in section \ref{ch:ThSumm}, where we list the key results, supporting them by leading considerations. The actual theory is presented in section \ref{ch:Theory}. Sections \ref{ch:ana} and \ref{ch:num} illustrate how to apply the method in practice. The summary of section \ref{ch:ThSumm} suffices for understanding the examples in sections \ref{ch:ana} and \ref{ch:num}. The reader who is not interested in a rigorous justification of the method can safely ignore it. Our first example is Kerr oscillator (section \ref{ch:ana}). This is an exactly soluble problem; moreover, all calculations in the truncated Wigner approach can be carried out analytically. In section \ref{ch:num} we apply the method to the Bose-Hubbard model, comparing the results to direct calculations in Hilbert space. We consider relatively small chains consisting of few sites to make exact calculations possible, although the truncated Wigner approach can be extended to much larger systems (see e.g. Refs.~\cite{ap_cat, adiabatic}). We find good agreement between exact and approximate results over relatively short time scales, with the applicability of this method to longer times remaining an open problem. 

\section{Theoretical summary}
\label{ch:ThSumm}
\subsection{Symmetric representation of Heisenberg operators}
\label{ch:TSH}
 
We will assume that the reader is familiar with the phase-space basics, including the concepts of symmetric (Weyl's) operator ordering, symmetric representation of operators and the Wigner function. The necessary minimum of information is summarised in section \ref{ch:PSB}. Details may be found in Refs.\ \cite{MandelWolf,walls-milburn,gardiner-zoller,AgarwalWolf}. 

The formal techniques we develop in this paper extend the well-known symmetric representation of quantum optics \cite{GardinerQN} to multitime problems. First and foremost, in place of a quasiaverage over the Wigner quasiprobability distribution we find a {\em phase-space path integral.\/} Within the path-integral techniques, we derive generalised {\em phase-space correspondenses\/} mapping multiplication of a q-number quantity by a creation or annihilation operator to phase-space. Conventional phase-space correspondenses apply to free operators and the generalised ones apply to Heisenberg operators. They are exact and do not depend on the nature of nonlinearity (interaction). Similarly to the way in which conventional phase-space correspondences allow one to reorder creation and annihilation operators, the generalised ones allow one to reorder a pair (say) of Heisenberg operators with unequal time arguments. 
These techniques are not restricted to any special Hamiltonian and can be employed for all bosonic systems.

To be specific, consider an illustrative example of the anharmonic oscillator (Kerr oscillator) with the Hamiltonian 
 {\begin{align}{{
 \begin{aligned} 
\hat{H}_0=\frac{\hbar\kappa}{2}\hat{a}^{\dagger 2}\hat{a}^2.
\end{aligned}}}%
\label{eq:69b} 
\end{align}}%
Here, $\hat{a},\hat{a}^{\dag}$ is the standard creation/annihilation pair, 
 {\begin{align}{{
 \begin{aligned} 
  \ensuremath{\big[
\hat{a},\hat{a}^{\dag}
\big]} = 1. 
\end{aligned}}}%
\label{eq:70b} 
\end{align}}%
The Heisenberg ``field operators'' are defined in the normal manner as 
 {\begin{gather}{{
 \begin{gathered} 
 {\hat{\mathcal A}}(t)= {\hat{\mathcal U}}^{\dagger}(t,t_0)\hat{a} {\hat{\mathcal U}}(t,t_0), \\  {\hat{\mathcal A}}^{\dagger}(t)=
 {\hat{\mathcal U}}^{\dagger}(t,t_0)\hat{a}^{\dagger} {\hat{\mathcal U}}(t,t_0). 
\end{gathered}}}%
\label{eq:71b} 
\end{gather}}%
Here, $ {\hat{\mathcal U}}(t,t_0)$ is the evolution operator, 
 {\begin{align}{{
 \begin{aligned} 
 {\hat{\mathcal U}}(t,t_0) = \exp  \ensuremath{\bigg[
-\frac{i(t-t_0)\hat{H}_0}{\hbar } 
\bigg]} , 
\end{aligned}}}%
\label{eq:72b} 
\end{align}}%
where $t_0$ is the coincidence point for the {Schr\"odinger}\ and {Heisenberg}\ pictures. 
For an arbitrary $\hat{H}_0$ the evolution operator is introduced through the {Schr\"odinger}\ equation, 
 {\begin{align}{{
 \begin{aligned} 
i\hbar \frac{\partial  {\hat{\mathcal U}}(t,t_0)}{\partial t} = 
\hat{H}_0 {\hat{\mathcal U}}(t,t_0), \ \  {\hat{\mathcal U}}(t_0,t_0) = \hat\openone. 
\end{aligned}}}%
\label{eq:73b} 
\end{align}}%

Note that we do not divide the Hamiltonian into the free and interaction part, nor introduce the interaction picture, nor free-field operators. The methods we develop in this paper are strictly nonperturbative. The truncated Wigner representation does not correspond to linearisation of any kind. It follows separation of nonlinearity and noise for path-integral trajectories with neglecting the latter, for details see section \ref{ch:TrW}. There does not seem to be a way of even formulating such approximations in Hilbert-space terms. 

In section \ref{ch:Theory} we construct a phase-space path-integral approach which allows one to calculate the quantum average of the symmetrised product 
 {\begin{align}{{
 \begin{aligned} 
  \ensuremath{\big\langle 
{\cal T}_W\!  {\hat{\mathcal A}}^{\dag}(t_2)
 {\hat{\mathcal A}}(t_1)
\big\rangle} = \overline{\overline{\hspace{0.1ex}\alpha (t_1)\alpha ^*(t_2)\hspace{0.1ex}}} , 
\end{aligned}}}%
\label{eq:58b} 
\end{align}}%
where 
 {\begin{align}{{
 \begin{aligned} 
{\cal T}_W\!  {\hat{\mathcal A}}^{\dag}(t_2)
 {\hat{\mathcal A}}(t_1) = \frac{1}{2}   \ensuremath{\Big[
 {\hat{\mathcal A}}^{\dag}(t_2)
 {\hat{\mathcal A}}(t_1) + 
 {\hat{\mathcal A}}(t_1) {\hat{\mathcal A}}^{\dag}(t_2) 
\Big]}. 
\end{aligned}}}%
\label{eq:75b} 
\end{align}}%
For the time being, ${\cal T}_W\! $ is just an {\em ad hoc\/} notation; its actual meaning will be the subject of section \ref{ch:TSO}. 
The quantum averaging is over the {Heisenberg}\ $\rho $-matrix $\hat \rho $, (with $ {\hat{\mathcal X}}$ being an arbitrary operator)
 {\begin{align}{{
 \begin{aligned} 
  \ensuremath{\big\langle 
 {\hat{\mathcal X}}
\big\rangle} = \text{Tr} \hat \rho  {\hat{\mathcal X}} . 
\end{aligned}}}%
\label{eq:76b} 
\end{align}}%
The double bar denotes symbolically the path integral over the c-number trajectories $\alpha (t)$. It can be thought of as a stochastic average with a nonpositive ``measure'', whose exact meaning will be clarified in the next section. What is important is that the path integral (\ref{eq:58b}) is {\em approximated\/} by the truncated Wigner approach. In this approximation the trajectories $\alpha (t)$ are deterministic and obey the equation
 {\begin{align}{{
 \begin{aligned}
i\frac{d\alpha(t)}{dt}=\kappa[|\alpha(t)|^2-1]\alpha(t) . 
\end{aligned}}}%
\label{eq:5B} 
\end{align}}%
This equation follows both from the traditional phase-space methods \cite{Wminus} and from the path-integral techniques in section \ref{ch:Theory}. The trajectories being deterministic, stochasticity in Eq.\ ( \ref{eq:58b}) reduces to the averaging over the Wigner function $W(\alpha)$ corresponding to the Heisenberg $\rho $-matrix (assuming this function is nonnegative, $W(\alpha )\geq 0$). The path integral in Eq.\ ( \ref{eq:58b}) then reduces to averaging over solutions of Eq.\ ( \ref{eq:5B}) with a random initial condition, 
 {\begin{align}{{
 \begin{aligned} 
\overline{\hspace{0.1ex}\alpha (t_1)\alpha ^*(t_2)\hspace{0.1ex}} = \int \frac{d^2\alpha }{\pi } W(\alpha) \alpha (t_1)\alpha ^*(t_2), 
\end{aligned}}}%
\label{eq:54b} 
\end{align}}%
where $\alpha (t)$ is a solution to (\ref{eq:5b}) with the initial condition $\alpha (t_0)=\alpha $. Making trajectories deterministic is an approximation, so that the average (\ref{eq:54b}) only approximates the quantum average, 
 {\begin{align}{{
 \begin{aligned} 
  \ensuremath{\big\langle 
{\cal T}_W\!  {\hat{\mathcal A}}^{\dag}(t_2)
 {\hat{\mathcal A}}(t_1)
\big\rangle} \approx \overline{\hspace{0.1ex}\alpha (t_1)\alpha ^*(t_2)\hspace{0.1ex}} , 
\end{aligned}}}%
\label{eq:87b} 
\end{align}}%
unlike the path integral (\ref{eq:58b}) which is exact. 

It is worth stressing here that the results of this paper are not Eqs.\ ( \ref{eq:5B})--(\ref{eq:87b}) as such, but the fact that they apply with $t_1\neq t_2$. The truncated Wigner approach was derived within the conventional phase-space techniques. By construction, it is only applicable to time-dependent averages of {Schr\"odinger}\ operators, or, equivalently, to averages of {Heisenberg}\ operators with equal time arguments. In other words, conventional phase-space methods allow one to verify Eqs.\ ( \ref{eq:5B})--(\ref{eq:87b}) only for $t_1=t_2$. Extension to unequal times (physically, to spectral properties of the system) requires alternative techniques such as a phase-space path integral. 
\subsection{Generalised phase-space correspondences}
\label{ch:TSG}
Our next goal is to extend the truncated-Wigner further by lifting the ordering restriction of Eq.\ ( \ref{eq:58b}). Namely, we wish to calculate the {\em time-normally ordered\/} average, 
{\begin{multline}\hspace{0.4\columnwidth}\hspace{-98.4pt} 
  \ensuremath{\big\langle
\hat {\cal A}^{\dagger}(t_1)\hat {\cal A}(t_2)
\big\rangle} =   \ensuremath{\big\langle 
{\cal T}_W\!  {\hat{\mathcal A}}^{\dag}(t_2)
 {\hat{\mathcal A}}(t_1)
\big\rangle} \\ 
- \frac{1}{2}  \ensuremath{\big\langle
  \ensuremath{\big[
\hat {\cal A}(t_2),\hat {\cal A}^{\dagger}(t_1)
\big]}
\big\rangle}
,
\hspace{0.4\columnwidth}\hspace{-98.4pt} 
\label{eq:2a} 
\end{multline}}%
where it has been expressed by the symmetrised average and the commutator. The symmetrised term is given by (\ref{eq:58b}). To express the commutator we employ the same idea as was used in Ref.\ \cite{QNDBWO} to reorder a time-normal average symmetrically. Namely, we assume that $t_2>t_1$ and relate the commutator to the linear response of the system, 
 {\begin{align}{{
 \begin{aligned} 
  \ensuremath{\big\langle   \ensuremath{\big[
 {\hat{\mathcal A}}(t_2), {\hat{\mathcal A}}^{\dag}(t_1)
\big]}
\big\rangle} = -i\hbar \frac{\delta   \ensuremath{\big\langle
 {\hat{\mathcal A}}(t_2)
\big\rangle} }{\delta s(t_1)} \bigg |_{s=0} .
\end{aligned}}}%
\label{eq:78b} 
\end{align}}%
This relation is simply Kubo's formula for the linear response function \cite{Kubo1,Kubo2} written ``from right to left.'' It implies that the Hamiltonian of the system has been complemented by an interaction with the external c-number source $s(t)$,
 {\begin{align}{{
 \begin{aligned}
\hat H_0 \to \hat H(t)= \hat H_0 - s(t)\hat a^{\dag} - s^*(t) \hat a .
\end{aligned}}}
\label{eq:79b} 
\end{align}}%
Strictly speaking, the condition $s=0$ must then be applied to both sides of Eq.\ ( \ref{eq:78b}) (and in fact to all quantum averages in the above), but in practice it suffices to remember that the only quantity defined with $s\neq 0$ is $  \ensuremath{\big\langle
 {\hat{\mathcal A}}(t_2)
\big\rangle} $ in Eq.\ ( \ref{eq:78b}). 

For simplicity we will confine the rest of the discussion to the truncated Wigner representation. In this case the trajectories $\alpha (t)$ are deterministic and obey the equation which differs from (\ref{eq:5B}) by the presence of an additive source:
 {\begin{align}{{
 \begin{aligned}
i\frac{d\alpha(t)}{dt}=\kappa[|\alpha(t)|^2-1]\alpha(t) - s(t). 
\end{aligned}}}%
\label{eq:5b} 
\end{align}}%
Within conventional phase-space methods, we find this equation by noting that the linear interaction terms in the Hamiltonian contribute only to drift terms in the generalised Fokker-Planck equation. A path-integral derivation of (\ref{eq:5b}) extending its applicability to multitime averages will be given in section \ref{ch:TrW}.

To calculate the linear response function (\ref{eq:79b}) one needs an infinitesimal instantaneous source at $t=t_1$, 
 {\begin{align}{{
 \begin{aligned}
s(t) = - i\hbar\, \delta \alpha\, \delta (t-t_1) .
\end{aligned}}}
\label{eq:80b}
\end{align}}%
Substituting this source into Eq.\ ( \ref{eq:5b}) we see that it causes a discontinuity of the trajectory at $t=t_1$ (a ``quantum jump'' in the terminology of Ref.\ \cite{ap_twa}). The additional factors in (\ref{eq:80b}) were chosen so as to make this discontinuity exactly equal to $\delta \alpha$. 
We thus have a simple correspondence
between sources and ``quantum jumps'':
 {\begin{align}{{
 \begin{aligned}
& \frac{\delta }{\delta s(t_1)} \Longleftrightarrow \frac{i}{\hbar } \frac{\partial }{\partial \alpha (t_1)},& 
\frac{\delta }{\delta s^*(t_1)} \Longleftrightarrow -\frac{i}{\hbar } \frac{\partial }{\partial \alpha^* (t_1)} .
\end{aligned}}}
\label{eq:Corr}
\end{align}}%
Mathematically, the derivatives ${\partial }/{\partial \alpha (t_1)},{\partial }/{\partial \alpha^* (t_1)}$ correspond to a variation of the {\em initial condition\/} set at $t=t_1$ instead of $t=t_0$. Such notation is to some extent informal but convenient. For the commutator we then find ($t_1<t_2$)
 {\begin{align}{{
 \begin{aligned}
  \ensuremath{\big\langle
  \ensuremath{\big[
\hat {\cal A}(t_2),\hat {\cal A}^{\dagger}(t_1)
\big]}
\big\rangle} =
\overline{\frac{\partial \alpha(t_2)}{\partial \alpha(t_1)}},\ \ t_2>t_1.
\end{aligned}}}%
\label{eq:4a} 
\end{align}}%
This relation follows from the correspondences (\ref{eq:Corr}) and from the path-integral representation of the average. 
 {\begin{align}{{
 \begin{aligned}
  \ensuremath{\big\langle
\hat {\cal A}(t_2)
\big\rangle} = \overline{\alpha (t_2)} .
\end{aligned}}}%
\label{eq:43b} 
\end{align}}%
Using (\ref{eq:4a}) we can express the time-normal average in the Wigner representation,
\begin{widetext}
 {\begin{align}{{
 \begin{aligned}
  \ensuremath{\big\langle
\hat {\cal A}^{\dagger}(t_1)\hat {\cal A}(t_2)
\big\rangle} \approx
\left\{
\begin{array}{ll}
\overline{\alpha^*(t_1)\alpha(t_2)-
\displaystyle\frac{1}{2} \frac{\partial {\alpha(t_2)}}{\partial \alpha(t_1)}},&t_1<t_2,
\\
\displaystyle\overline{\alpha^*(t_1)\alpha(t_2)-\frac{1}{2} \bigg[\frac{\partial {\alpha(t_1)}}{\partial \alpha(t_2)}\bigg]^*},&t_1>t_2\, .
\end{array}
\right .
\end{aligned}}}%
\label{eq:wichtig} 
\end{align}}%
\end{widetext}
The second line here follows by conjugating the first one and replacing $t_1\leftrightarrow t_2$. 

These {\em generalised phase-space correspondences\/} (\ref{eq:wichtig}) are the central result of the paper. Certainly, the above derivation is no more than leading considerations, but the rigorous treatment in section \ref{ch:RPh} gives the same result. Moreover, it shows that Eqs.\ ( \ref{eq:wichtig}) hold not only as an approximation in the truncated Wigner representation, but also as an exact relation within the rigorous path-integral approach --- in which case the bar in (\ref{eq:wichtig}) should be replaced by double bar. 

Why do we call Eqs.\ ( \ref{eq:wichtig}) ``generalised phase-space correspondences?'' To recognise the connection we take the limit $t_1,t_2\to t_0$. Using the fact that that $\hat {\cal A}(t_0)=\hat a, \hat {\cal A}^{\dagger}(t_0)=\hat a^{\dag}$ and dropping the time argument in $\alpha (t_0)$ we find 
 {\begin{align}{{
 \begin{aligned} 
  \ensuremath{\big\langle 
\hat a^{\dag} \hat a
\big\rangle} 
= \overline{\hspace{0.1ex} \ensuremath{\bigg(
\alpha - \frac{1}{2} \frac{\partial }{\partial \alpha ^*} 
 \bigg)}\alpha ^*\hspace{0.1ex}} 
= \overline{\hspace{0.1ex} \ensuremath{\bigg(
\alpha^* - \frac{1}{2} \frac{\partial }{\partial \alpha} 
 \bigg)}\alpha\hspace{0.1ex}} 
, 
\end{aligned}}}%
\label{eq:55b} 
\end{align}}%
where the averaging is simply over the Wigner function $W(\alpha)$, cf.\ Eq.\ ( \ref{eq:54b}). First of all, both relations in Eq.\ ( \ref{eq:55b}) are correct. Indeed, they result in the formula
 {\begin{align}{{
 \begin{aligned} 
  \ensuremath{\big\langle 
\hat a^{\dag} \hat a
\big\rangle} = \overline{\hspace{0.1ex}|\alpha |^2\hspace{0.1ex}} - \frac{1}{2}\, . 
\end{aligned}}}%
\label{eq:82b} 
\end{align}}%
The average over the Wigner function expresses symmetrically ordered products of $\hat a, \hat a^{\dag}$; in particular, 
 {\begin{align}{{
 \begin{aligned} 
\overline{\hspace{0.1ex}|\alpha |^2\hspace{0.1ex}} = \frac{1}{2}   \ensuremath{\big\langle 
\hat a \hat a^{\dag} + 
\hat a^{\dag} \hat a
\big\rangle} =   \ensuremath{\big\langle 
\hat a^{\dag} \hat a
\big\rangle} + \frac{1}{2}, 
\end{aligned}}}%
\label{eq:83b} 
\end{align}}%
in obvious agreement with (\ref{eq:82b}). Furthermore, Eqs.\ ( \ref{eq:55b}) are particular cases of the standard phase-space correspondences, 
\begin{gather} 
{\begin{aligned} 
  \ensuremath{\Big\langle 
{\hat Y \hat a}
\Big\rangle} = \overline{\hspace{0.1ex} \ensuremath{\bigg(
\alpha - \frac{1}{2} \frac{\partial }{\partial \alpha ^*} 
 \bigg)} Y(\alpha)\hspace{0.1ex}} , 
\end{aligned}}%
\label{eq:84b} 
\\ 
{\begin{aligned} 
  \ensuremath{\Big\langle 
{\hat a^{\dag}\hat Y }
\Big\rangle} = \overline{\hspace{0.1ex} \ensuremath{\bigg(
\alpha^* - \frac{1}{2} \frac{\partial }{\partial \alpha} 
 \bigg)} Y(\alpha)\hspace{0.1ex}} , 
\end{aligned}}%
\label{eq:85b} 
\end{gather}%
where $\hat Y$ is an operator and $Y(\alpha )$ is its symmetric representation. The first of Eqs.\ ( \ref{eq:55b}) follows from Eq.\ ( \ref{eq:84b}) with $\hat Y = \hat a^{\dag}, Y(\alpha ) = \alpha ^*$, and the second one --- from Eq.\ ( \ref{eq:85b}) with $\hat Y = \hat a, Y(\alpha ) = \alpha$. Multitime generalisations of Eqs.\ ( \ref{eq:84b}), (\ref{eq:85b}) are derived in section \ref{ch:RPh}. 

We conclude this paragraph with a remark on terminology. To maintain rigor, one should distinguish {\em shifts\/} of trajectories effected by instantaneous sources (\ref{eq:80b}) from {\em quantum jumps\/}. The latter term was introduced in Ref.\ \cite{ap_twa}, where discontinuities of trajectories were used as formal means to express perturbative corrections to the truncated Wigner approach. These corrections come from quantum noises which do not have any classical interpretation whatsoever, while the c-number external source is to a large extent a classical object. Maintaining the distinction between shifts and quantum jumps thus appears physically justified. However, such clear-cut distinction is an artifact of an undamped model with quartic interaction. For instance, for the damped harmonic oscillator the equation for the Wigner function is a genuine Fokker-Planck equation. In this case the ``quantum noise'' is fully probabilistic, i.e., classical. We will use ``quantum jump'' as a blanket term applicable to both types of discontinuities. 

\subsection{The time-symmetric ordering}
\label{ch:TSO}
The fact that the path integral (\ref{eq:58b}) calculates (and the truncated Wigner approach approximates) symmetrised products of Heisenberg operators does not generalise to products of three and more operators. Instead of fully symmetrised products, one discovers a new type of ordering of Heisenberg operators, which we call {\em time-symmetric\/} and denote as ${\cal T}_W\!$. We find this interesting and important enough to be worth reporting, notwithstanding the fact that it is not directly relevant to purposes of this paper. 

A time-symmetrically ordered product of the ``field operators'' $ {\hat{\mathcal A}}(t), {\hat{\mathcal A}}^{\dag}(t)$ is defined recursively as 
 {\begin{align}{{
 \begin{aligned} 
& {\cal T}_W\! \hat\openone = \hat\openone , \ \ 
{\cal T}_W\!  {\hat{\mathcal A}}(t) =  {\hat{\mathcal A}}(t), \ \ 
{\cal T}_W\!  {\hat{\mathcal A}}^{\dag}(t) =  {\hat{\mathcal A}}^{\dag}(t) , \\
& {\cal T}_W\!  {\hat{\mathcal A}}(t) {\hat{\mathcal P}}_{[>t]} = \frac{1}{2}   \ensuremath{\big\{
 {\hat{\mathcal A}}(t),{\cal T}_W\!  {\hat{\mathcal P}}_{[>t]} 
\big\}} , \\ 
& {\cal T}_W\!  {\hat{\mathcal A}}^{\dag}(t) {\hat{\mathcal P}}_{[>t]} = \frac{1}{2}   \ensuremath{\big\{
 {\hat{\mathcal A}}^{\dag}(t),{\cal T}_W\!  {\hat{\mathcal P}}_{[>t]} 
\big\}} . 
\end{aligned}}}%
\label{eq:46b} 
\end{align}}%
Here, $ {\hat{\mathcal P}}_{[>t]}$ is a product of field operators with all time arguments exceeding $t$; the curly brackets stand for the anticommutator, $  \ensuremath{ \{
 {\hat{\mathcal X}}, {\hat{\mathcal Y}}
 \}} =  {\hat{\mathcal X}} {\hat{\mathcal Y}}+ {\hat{\mathcal Y}} {\hat{\mathcal X}}$. It is implied that under the sign of ${\cal T}_W\! $-ordering the field operators commute freely. The quantum average of an arbitrary time-symmetric product is expressed as a path-integral average, ($m,n\geq 0$) 
{\begin{multline}\hspace{0.4\columnwidth}\hspace{-98.4pt} 
  \ensuremath{\big\langle 
{\cal T}_W\! {\hat{\mathcal A}}(t_1)\cdots
 {\hat{\mathcal A}}(t_m)
 {\hat{\mathcal A}}^{\dag}(t'_1)\cdots
 {\hat{\mathcal A}}^{\dag}(t'_n)
\big\rangle} \\ = \overline{\overline{\hspace{0.1ex}\alpha(t_1)\cdots
\alpha(t_m)
\alpha^*(t'_1)\cdots
\alpha^*(t'_n)
\hspace{0.1ex}}} .
\hspace{0.4\columnwidth}\hspace{-98.4pt} 
\label{eq:86b} 
\end{multline}}%
For the exact meaning of this relation we refer the reader to the section \ref{ch:Theory}. Again, what matters is that the truncated Wigner approach represents the path-integral average approximately, 
 {\begin{align}{{
 \begin{aligned} 
&   \ensuremath{\big\langle 
{\cal T}_W\! {\hat{\mathcal A}}(t_1)\cdots
 {\hat{\mathcal A}}(t_m)
 {\hat{\mathcal A}}^{\dag}(t'_1)\cdots
 {\hat{\mathcal A}}^{\dag}(t'_n)
\big\rangle} \\ & \ \ \ \approx \overline{\hspace{0.1ex}\alpha(t_1)\cdots
\alpha(t_m)
\alpha^*(t'_1)\cdots
\alpha^*(t'_n)
\hspace{0.1ex}} \\ & \ \ \ = \int \frac{d^2\alpha }{\pi } W(\alpha) 
\alpha(t_1)\cdots
\alpha(t_m)
\alpha^*(t'_1)\cdots
\alpha^*(t'_n)
, 
\end{aligned}}}%
\label{eq:88b} 
\end{align}}%
cf.\ Eq.\ ( \ref{eq:54b}). Equations (\ref{eq:46b}) and (\ref{eq:86b}) may be directly generalised to multimode and real-space cases, by supplementing the time arguments by suitable ``labels,'' such as mode indices or spatial arguments. For an example (the Bose-Hubbard chain) see section \ref{ch:num}. 

The two most important properties of the time-symmetric products are: these products are continuous at coinciding time arguments, and for free-field operators they turn into the conventional symmetric (Weyl) ordered products. For two operators, the time-symmetric product coincides with a symmetrised product given by Eq.\ ( \ref{eq:75b}). 
That quantity is continuous at $t=t'$; moreover, for coinciding times, (\ref{eq:75b}) reduces to the conventional formula for the symmetric ordering, which naturally appears in the --W approximation~\cite{walls-milburn,gardiner-zoller,blakie},
 {\begin{align}{{
 \begin{aligned} 
\textrm{W}  \ensuremath{\big\{
\hat a \hat a ^{\dag}
\big\}} = 
\frac{1}{2}  \ensuremath{\big(
\hat a \hat a ^{\dag} + 
\hat a^{\dag} \hat a 
 \big)} . 
\end{aligned}}}%
\label{eq:47b} 
\end{align}}%
(Recall that for coinciding times the field operators commute the same way as the creation and annihilation operators.)
For three operators and $t_1<t_2<t_3$ we have, for example, 
{\begin{multline}\hspace{0.4\columnwidth}\hspace{-98.4pt} 
{\cal T}_W  {\hat{\mathcal A}}(t_1) {\hat{\mathcal A}}(t_2) {\hat{\mathcal A}}^{\dag}(t_3) \\ = \frac{1}{4}   \ensuremath{\big[
 {\hat{\mathcal A}}(t_1) {\hat{\mathcal A}}(t_2) {\hat{\mathcal A}}^{\dag}(t_3)+ 
 {\hat{\mathcal A}}(t_2) {\hat{\mathcal A}}^{\dag}(t_3) {\hat{\mathcal A}}(t_1) \\ +
 {\hat{\mathcal A}}(t_1) {\hat{\mathcal A}}^{\dag}(t_3) {\hat{\mathcal A}}(t_2) +
 {\hat{\mathcal A}}^{\dag}(t_3) {\hat{\mathcal A}}(t_2) {\hat{\mathcal A}}(t_1)
\big]} .
\hspace{0.4\columnwidth}\hspace{-98.4pt} 
\label{eq:49b}
\end{multline}}%
Here the time-symmetric ordering is not the same as the fully symmetric ordering; in the latter there should be two additional terms with $ {\hat{\mathcal A}}(t_1)$ in the middle. Again, it may be shown that (\ref{eq:49b}) is continuous at coinciding time arguments, and that with all three times equal it agrees with the formula for the Weyl-ordered product, 
 {\begin{align}{{
 \begin{aligned} 
\textrm{W}  \ensuremath{\big\{
\hat a^2\hat a^{\dag} 
\big\}} = \frac{1}{3}\left(
\hat a^2 \hat a^{\dag} + \hat a^{\dag}\hat a^2 + \hat a \hat a^{\dag} \hat a
\right). 
\end{aligned}}}%
\label{eq:50b} 
\end{align}}%
Detailed discussion of the time-symmetric ordering requires advanced formal tools and will be presented elsewhere \cite{Unpb}. 

We note that all operator products entering the time-symmetric product exhibit a special order of time arguments: times first increase then decrease. Such order of operators is characteristic of Schwinger's closed-time-loop formalism \cite{SchwingerC}. This connection is investigated in Ref.\ \cite{Unpb}. 
We also note without proof that only such ``Schwinger-ordered'' operator products have causal representation through quantum jumps similar to Eqs.\ ( \ref{eq:wichtig}). This restriction becomes nontrivial for products of three or more operators. For example there is no causal representation through the response for finding the expectation value of $ {\hat{\mathcal A}}(t_2) {\hat{\mathcal A}}(t_1) {\hat{\mathcal A}}(t_3)$ with $t_1<t_2,t_3$. For this particular ordering one cannot avoid finding the response at $t_1$ to a perturbation which happens later in the evolution either at $t=t_2$ or $t=t_3$. For more details see Ref.~\cite{ap_wig}. 
\section{Multitime Wigner approach}
\label{ch:Theory}

\subsection{Preliminary remarks}\label{ch:PR}
 
In this section we present a rigorous derivation of the ``generalised phase-space correspondences'' (\ref{eq:wichtig}). The reader who is interested only in applications of the method can safely skip the formalism and go directly to examples in sections \ref{ch:ana} and \ref{ch:num}. 

For simplicity we will continue working with the illustrative example of the Kerr oscillator. The necessary definitions were given in section \ref{ch:TSH}. In fact all formulae in this section apply to arbitrary time-dependent Hamiltonians, and can also be easily generalised to multimode problems, simply by complementing the time arguments by other ``labels,'' such as mode indices or spatial arguments. 
\subsection{Phase-space basics}\label{ch:PSB}
For the reader's convenience, we summarise here the necessary facts from phase-space techniques \cite{MandelWolf,walls-milburn,gardiner-zoller,AgarwalWolf}. The displacement operator is defined as 
 {\begin{align}{{
 \begin{aligned}
\hat D(\alpha ) = \text{e}^{\alpha \hat a^{\dag} - \alpha ^* \hat a } .
\end{aligned}}}
\label{eq:D} 
\end{align}}%
For an arbitrary operator $\hat A$, one introduces its characteristic function,
 {\begin{align}{{
 \begin{aligned}
\chi _A(\alpha ) = \text{Tr} \hat A \hat D^{\dag}(\alpha ),
\end{aligned}}}
\label{eq:ChF} 
\end{align}}%
and its symmetric representation,
 {\begin{align}{{
 \begin{aligned}
A(\alpha)=  \ensuremath{\big[
\hat{A}
\big]} (\alpha)=\int\frac{d^2\beta}{\pi}
\chi _A(\beta )
e^{\beta\alpha^*-\beta^*\alpha}
.
\end{aligned}}}
\label{eq:SR} 
\end{align}}%
Expressions for $\hat A$ in terms of these read
 {\begin{align}{{
 \begin{aligned}
\hat{A}&=
\int \frac{d^2\beta}{\pi}\hat{D}(\beta)\chi_A (\beta)
\\ &=
\int \frac{d^2\alpha d^2\beta}{\pi^2}e^{\alpha \beta^*-\alpha^*\beta}A(\alpha)\hat{D}(\beta) \, .
\end{aligned}}}
\label{eq:DefW} 
\end{align}}%
The notation $[\cdots](\alpha )$ is convenient for symmetric representations of operator expressions, as, for instance, in Eqs.\ ( \ref{eq:AB1}), (\ref{eq:AB2}) below (see also endnote \cite{EndSB}). 
{
Of use to us will be the relations, 
 {\begin{align}{{
 \begin{aligned} 
&   \ensuremath{\big[
\hat a
\big]}(\alpha ) = \alpha , \ \ 
  \ensuremath{\big[
\hat a^{\dag}
\big]}(\alpha ) = \alpha^* , \\
&   \ensuremath{\big[
\hat a^{\dag}\hat a
\big]}(\alpha ) = |\alpha|^2 - \frac{1}{2} , \\
&   \ensuremath{\big[
\hat a^{\dag 2}\hat a^2
\big]}(\alpha ) = |\alpha|^4 - 2|\alpha|^2 + \frac{1}{2} . 
\end{aligned}}}%
\label{eq:90b} 
\end{align}}%
}

Displacement operators form a complete set with respect to the Hilbert-Schmidt norm,
{\begin{multline}\hspace{0.4\columnwidth}\hspace{-98.4pt} 
\text{Tr} \hat A \hat B = \int \frac{d^2\alpha }{\pi }
\,\text{Tr} \hat A\hat D(\alpha )\,
\text{Tr}\hat D^{\dag}(\alpha ) \hat B \\ =
\int \frac{d^2\alpha }{\pi }
A(\alpha )B(\alpha )\, . 
\hspace{0.4\columnwidth}\hspace{-98.4pt} 
\label{eq:TrW} 
\end{multline}}%
The last equation here is a consequence of (\ref{eq:SR}). In particular, it allows one to write a phase-space representation of a quantum average,
 {\begin{align}{{
 \begin{aligned}
  \ensuremath{\big\langle
\hat A
\big\rangle} = \text{Tr}\hat A \hat \rho = \int \frac{d^2 \alpha }{\pi }A(\alpha )\rho (\alpha ) .
\end{aligned}}}
\label{eq:AvW} 
\end{align}}%

Of importance to us will be a relation expressing the Wigner representation of an operator product $\hat A \hat B$ by the Wigner representations of the factors:
{\begin{multline}\hspace{0.4\columnwidth}\hspace{-98.4pt} 
{[\hat A\hat B]}(\alpha) = \int \frac{d^2 \alpha _0 d^2\sigma }{\pi ^2}
\text{e}^{(\alpha -\alpha _0)\sigma ^* - (\alpha -\alpha _0)^*\sigma } \\ \times A(\alpha_0 )
B(\alpha_0 + \sigma /2) .
\hspace{0.4\columnwidth}\hspace{-98.4pt} 
\label{eq:AB1}
\end{multline}}%
By the change of variable $\alpha _0\to\alpha _0 + \sigma /2$ we can write this in the alternative form,
{\begin{multline}\hspace{0.4\columnwidth}\hspace{-98.4pt} 
{[\hat A\hat B]}(\alpha) = \int \frac{d^2 \alpha _0 d^2\sigma }{\pi ^2}
\text{e}^{(\alpha -\alpha _0)\sigma ^* - (\alpha -\alpha _0)^*\sigma } \\ \times A(\alpha_0 - \sigma /2)
B(\alpha_0 ) .
\hspace{0.4\columnwidth}\hspace{-98.4pt} 
\label{eq:AB2}
\end{multline}}%
These relations follow from expressing the operators by their symmetric representations using~Eq.\ ( \ref{eq:DefW}) and then employing~Eq.\ ( \ref{eq:SR}) to express $[\hat A\hat B](\alpha)$. It is easy to verify that \cite{AgarwalWolf}
{\begin{multline}\hspace{0.4\columnwidth}\hspace{-98.4pt} 
\text{Tr}\hat D(\alpha )\hat D(\beta ) \hat D(\beta ') \\ =
\pi \delta ^{(2)}(\alpha +\beta +\beta ')
\text{e}^{\frac{1}{2}(\beta \beta ^{\prime *} - \beta^* \beta' ) } .
\hspace{0.4\columnwidth}\hspace{-98.4pt} 
\end{multline}}%
The rest of the calculation leading to Eqs.\ ( \ref{eq:AB1}), (\ref{eq:AB2}) is straightforward.

The symmetric representation of an operator is often introduced as an expression for this operator in terms of symmetrically (Weyl) ordered products of creation and annihilation operators. Such products are defined postulating that 
 {\begin{align}{{
 \begin{aligned} 
  \ensuremath{\big[
\textrm{W}  \ensuremath{\big\{
\hat a^m \hat a^{\dag n}
\big\}} 
\big]}(\alpha ) = \alpha ^m\alpha ^{* n} . 
\end{aligned}}}%
\label{eq:91b} 
\end{align}}%
Equations (\ref{eq:90b}) are then written as operator formulae, 
 {\begin{align}{{
 \begin{aligned} 
& \hat a = \textrm{W}  \ensuremath{\big\{
\hat a 
\big\}} , \ \ 
\hat a^{\dag} = \textrm{W}  \ensuremath{\big\{
\hat a^{\dag}
\big\}} , \\
&
\hat a^{\dag}\hat a = \textrm{W}  \ensuremath{\big\{
\hat a \hat a^{\dag}
\big\}} - \frac{1}{2} , \\
& 
\hat a^{\dag 2}\hat a^2 = \textrm{W}  \ensuremath{\big\{
\hat a^2 \hat a^{\dag 2}
\big\}} - 2\textrm{W}  \ensuremath{\big\{
\hat a\hat a^{\dag}
\big\}} + \frac{1}{2} . 
\end{aligned}}}%
\label{eq:93b} 
\end{align}}%
These equations (and thus Eqs.\ ( \ref{eq:90b})) may be verified noting that the displacement operator is naturally Weyl-ordered, 
 {\begin{align}{{
 \begin{aligned} 
\hat D(\alpha ) = \textrm{W}  \ensuremath{\big\{\hat D(\alpha )\big\}} , 
\end{aligned}}}%
\label{eq:92b} 
\end{align}}%
and can therefore serve as an operator-valued characteristic function for the symmetrically ordered products, 
 {\begin{align}{{
 \begin{aligned} 
\hat D(\alpha ) = \sum_{m,n=0}^{\infty} \frac{\alpha ^n(-\alpha ^{*})^m}{m!n!}\, \textrm{W}  \ensuremath{\big\{
\hat a^m \hat a^{\dag n}
\big\}} . 
\end{aligned}}}%
\label{eq:94b} 
\end{align}}%
Verification of Eqs.\ ( \ref{eq:93b}) reduces to developing $\hat D(\alpha )$ in a power series, with the subsequent use of (\ref{eq:70b}). 
\subsection{Phase-space transition amplitude}
With the only exception of Eq.\ ( \ref{eq:D}) which employs the creation/annihilation pair in the Schr\"odinger picture, all definitions in section \ref{ch:PSB} may be applied to Schr\"odinger as well as to Heisenberg operators. If a particular operator is time-dependent, its symmetric representation is also time-dependent. 
The time-dependent symmetric representation of a Schr\"odinger operator and the time-dependent Wigner function are defined as follows,
 {\begin{align}{{
 \begin{aligned}
  \ensuremath{\big[
\hat B(t)
\big]} (\alpha ) = B(\alpha ,t), \ \ \
  \ensuremath{\big[
\hat \rho (t)
\big]} (\alpha ) = \rho (\alpha ,t),
\end{aligned}}}
\end{align}}%
cf.\ Eq.\ ( \ref{eq:SR}). We stress that both definitions here are for operators in the Schr\"odinger picture. In the Heisenberg picture the density matrix is stationary and coincides with $\hat\rho(t_0)$; its symmetric representation thus coincides with $\rho(\alpha,t_0)$. The Heisenberg counterpart of $\hat B(t)$ reads 
 {\begin{align}{{
 \begin{aligned} 
 {\hat{\mathcal B}}(t) =  {\hat{\mathcal U}}^{\dag}(t,t_0)\hat B(t) {\hat{\mathcal U}}(t,t_0) . 
\end{aligned}}}%
\label{eq:95b} 
\end{align}}%
We do not introduce any special notation for symmetric representations of {Heisenberg}\ operators but use the bracket symbol instead, cf.\ Eq.\ ( \ref{eq:Bt}) below. 

Using Eq.\ ( \ref{eq:TrW}), the quantum average of $ {\hat{\mathcal B}}(t)$ may be written as
 {\begin{align}{{
 \begin{aligned}
  \ensuremath{\big\langle
 {\hat{\mathcal B}}(t)
\big\rangle} = \int \frac{d^2 \alpha }{\pi }B(\alpha ,t)
  \ensuremath{\big[
 {\hat{\mathcal U}}(t,t_0)\rho (t_0) {\hat{\mathcal U}}^{\dag}(t,t_0)
\big]} (\alpha)
\\ =\int\frac{d^2\alpha_0d^2\alpha}{\pi^2}
B(\alpha,t)U(\alpha,t,\alpha_0,t_0)\rho(\alpha_0,t_0),
\end{aligned}}}
\label{eq:BAv} 
\end{align}}%
where we have introduced the {\em phase-space transition amplitude\/}
{\begin{multline}\hspace{0.4\columnwidth}\hspace{-98.4pt} 
U(\alpha,t,\alpha_0,t_0) = \int\frac{d^2\beta_0d^2\beta}{\pi^2}e^{\alpha\beta^*-\alpha^*\beta+\alpha_0\beta_0^*-\alpha_0^*\beta_0}
\\ \times
\text{Tr}\hat{D}^{\dagger}(\beta) \hat{U}(t,t_0)\hat{D}^{\dagger}(\beta_0) \hat{U}^{\dagger}(t,t_0)\, .
\hspace{0.4\columnwidth}\hspace{-98.4pt} 
\label{eq:ta} 
\end{multline}}%
By construction, this amplitude evolves the Wigner function in time,
\begin{align}
\rho(\alpha,t)=\int\frac{d^2\alpha_0}{\pi}U(\alpha,t,\alpha_0,t_0)\rho(\alpha_0,t_0)\, ,
\label{eq:rho}
\end{align}
but it can also be applied to the operator,
\begin{align}
[\hat{\mathcal{B}}_{t_0}(t)](\alpha_0)=\int\frac{d^2\alpha}{\pi}B(\alpha,t)U(\alpha,t,\alpha_0,t_0)\, .
\label{eq:Bt}
\end{align}
In this formula the dependence of the Heisenberg operator on the coincidence point $t_0$ is made explicit showing it as a subscript. Such notation is convenient when the coincidence point itself becomes a variable as in section \ref{ch:MultT} below. 
\subsection{Phase-space path integral and the truncated Wigner representation}\label{ch:TrW}
The group property of the evolution operator 
\begin{align}
\hat{U}(t,t_0)= \hat{U}(t,t_1)\hat{U}(t_1,t_0), 
\ \ t>t_1>t_0, 
\end{align}
results in the related property of the transition amplitude,
\begin{align}
U(\alpha, t, \alpha_0, t_0) = \int\frac{d^2\alpha_1}{\pi}U(\alpha,t,\alpha_1,t_1)
U(\alpha_1,t_1,\alpha_0,t_0)\,.
\label{eq:group}
\end{align}
Breaking the time interval $[t_0,t]$ into $M+1$ Trotter slices,
 {\begin{align}{{
 \begin{aligned}
\Delta t=\frac{t-t_0}{M+1}, \ \ \
t_k=t_0+k\Delta t, ~ k=0,\ldots,M,
\end{aligned}}}
\end{align}}%
we can define the {\em path-integral representation\/} of the phase-space amplitude as the limit
{\begin{multline}\hspace{0.4\columnwidth}\hspace{-98.4pt} 
U(\alpha, t, \alpha_0, t_0) =
\lim_{M\rightarrow\infty}\int U(\alpha,t,\alpha_M,t_M)
\\ \times
\prod_{k=1}^M\frac{d^2\alpha_k}{\pi}U(\alpha_k,t_k,\alpha_{k-1},t_{k-1}) \, .
\hspace{0.4\columnwidth}\hspace{-98.4pt} 
\label{eq:takz} 
\end{multline}}%
Each amplitude on the RHS here is over an infinitesimal time interval $\Delta t$.

To understand the path integral we have thus to understand the infinitesimal transition amplitude. It may be evaluated using the method introduced by one of us in Ref.~\cite{ap_twa}. We start from the von-Neuman equation for the density matrix
\begin{align}
i\hbar \dot{\hat{\rho}}(t)=[\hat{H}(t),\hat{\rho}(t)],
\end{align}
so that
 {\begin{align}{{
 \begin{aligned}
\hat\rho (t+\Delta t) = \hat\rho (t)
- \frac{i\Delta t}{\hbar } \hat{H}(t)\hat\rho (t)
+ \frac{i\Delta t}{\hbar } \hat\rho (t)\hat{H}(t).
\end{aligned}}}%
\label{eq:25b} 
\end{align}}%

Note that we wrote $\hat H(t)$ to highlight the fact that the derivation is valid for arbitrary time-dependent Hamiltonians, including Hamiltonians with external sources such as (\ref{eq:79b}). 
Employing Eqs.\ ( \ref{eq:AB1}), (\ref{eq:AB2}) and introducing the symmetric representation of the Hamiltonian in the Schr\"odinger picture,
 {\begin{align}{{
 \begin{aligned}
H(\alpha ,t) = [\hat H(t)](\alpha ),
\end{aligned}}}%
\label{eq:27b} 
\end{align}}%
we have
\begin{widetext}
 {\begin{align}{{
 \begin{aligned}
\rho(\alpha,t+\Delta t) = \int \frac{d^2\alpha_0d^2\sigma}{\pi^2}
e^{(\alpha-\alpha_0)\sigma^*-(\alpha-\alpha_0)^*\sigma}
\rho(\alpha_0,t)
  \ensuremath{\Bigg\{
1-\frac{i\Delta t}{\hbar}   \ensuremath{\Bigg[
H \ensuremath{\Bigg(
\alpha_0-\frac{\sigma}{2},t
 \Bigg)}-H \ensuremath{\Bigg(\alpha_0+\frac{\sigma}{2},t \Bigg)}
\Bigg]}
\Bigg\}} ,
\end{aligned}}}%
\label{eq:VGL} 
\end{align}}%
see also endnote \cite{EndSB}. Comparing this to Eq.\ ( \ref{eq:rho}) and using the fact that $\Delta t$ is infinitesimally small we find
{\begin{multline}
 U(\alpha,t+\Delta t, \alpha_0, t)
\\
= \int\frac{d^2\sigma}{\pi} \exp  \ensuremath{\bigg\{
  \ensuremath{\bigg[
\alpha-\alpha_0+if(\alpha_0, t)\frac{\Delta t}{\hbar}
\bigg]} \sigma^*
-
  \ensuremath{\bigg[
\alpha-\alpha_0+if(\alpha_0, t)\frac{\Delta t}{\hbar}
\bigg]} ^*\sigma
+\frac{i\Delta t}{\hbar}h^{(3)}(\alpha_0,\sigma,t)
\bigg\}} \, ,
\label{eq:42b} 
\end{multline}}%
\end{widetext}%
where $f(\alpha_0, t)$ and $h^{(3)}(\alpha_0,\sigma,t)$ are found by expanding the symmetric representation of the interaction Hamiltonian into power series:
 {\begin{align}{{
 \begin{aligned}
& f(\alpha_0, t)=\frac{\partial H(\alpha_0,t)}{\partial \alpha_0^*},
\ \ \
f^*(\alpha_0, t)=\frac{\partial H(\alpha_0,t)}{\partial \alpha_0},
\\
& {\begin{aligned}
h^{(3)}(\alpha_0,\sigma,t) & = H\bigg(\alpha_0+\frac{\sigma}{2},t\bigg)-H\bigg(\alpha_0-\frac{\sigma}{2},t\bigg)
\\ & \ \ \
- \sigma f^*(\alpha_0, t) - \sigma^*f(\alpha_0,t)\, .
\end{aligned}}
\end{aligned}}}
\label{eq:fh} 
\end{align}}%
The term $h^{(3)}(\alpha_0,\sigma,t)$ is responsible for {\em cubic noise\/}, which accounts for quantum fluctuations; a consistent derivation of the path integral with the cubic noise will be subject of a separate paper.
Attempts to simulate the cubic noise numerically were rather disappointing \cite{nossoEPL,ourpreprint,spontaneous}. In Ref.~\cite{ap_twa} one of us showed how it can be taken into account perturbatively through the nonlinear response. In Ref.~\cite{adiabatic} this nonlinear response was implemented to improve the accuracy of the --W for a large BH chain of 128 sites.

On the other hand, neglecting cubic noises simplifies our task enormously by removing all mathematical problems associated with their highly singular nature. Without $h^{(3)}$ the integral in (\ref{eq:42b}) is calculated straightaway, and we have
\begin{align}
U(\alpha,t+\Delta t,\alpha_0,t_0)=\pi\delta^{(2)} \ensuremath{\bigg(
\alpha-\alpha_0+\frac{i}{\hbar}f(\alpha_0, t)\Delta t
 \bigg)} \, .
\end{align}
This corresponds to a deterministic evolution in phase-space along the trajectories satisfying the equation
\begin{align}
i\hbar \dot{\alpha}=f(\alpha,t)\, .
\label{eq:GtW}
\end{align}
By making use of Eqs.\ ( \ref{eq:90b}), for the Kerr oscillator we find Eq.\ ( \ref{eq:5B}). We have thus recovered the well-known {\em truncated Wigner representation\/} \cite{Wminus}. However, unlike in Ref.\ \cite{Wminus}, we have found it as an approximation within a consistent phase-space path-integral approach. This allows us to answer two questions which cannot be answered in the derivation based on the Fokker-Planck equation. Firstly, which quantum averages the path integral calculates, and, secondly, how one could evaluate other types of averages. This will be subject of Secs.\
\ref{ch:MultT} and \ref{ch:RPh}.

Equations (\ref{eq:fh}) make it obvious that external sources in the Hamiltonian manifest themselves as additive sources in the equations for trajectories. Indeed, using Eqs.\ ( \ref{eq:90b}), for the Hamitonian (\ref{eq:79b}) we have, 
 {\begin{align}{{
 \begin{aligned} 
  \ensuremath{\big[
\hat H(t)
\big]}(\alpha ) = 
  \ensuremath{\big[
\hat H_0\big]}(\alpha ) - s(t)\alpha^* - s^*(t) \alpha. \end{aligned}}}%
\label{eq:96b} 
\end{align}}%
The source terms only modify the regular evolution,
 {\begin{gather}{{
 \begin{gathered}
f(\alpha ,t) \to f(\alpha ,t) - s(t), \\ f^*(\alpha ,t) \to f^*(\alpha ,t) - s^*(t),
\end{gathered}}}
\label{eq:ToS}
\end{gather}}%
For the Kerr oscillator this results in Eqs.\ ( \ref{eq:5b}). Equation (\ref{eq:96b}) holds for an arbitrary $\hat H_0$, so that replacements (\ref{eq:ToS}) apply in general. 

That the Kubo-style sources in the Hamiltonian appear as additive sources in the equations of motion for the phase-space trajectories is in fact true for arbitrary phase-space techniques. 
Indeed, irrespective of the operator ordering, linear terms in the Hamiltonian manifest themselves {\em only\/} as drift terms in the generalised Fokker-Planck equation and thus {\em only\/} as additive terms in the corresponding generalised Langevin equations. For an example see Ref.\ \cite{QNDBWO}, where external sources were introduced in the positive-P representation. 
\subsection{Time-symmetric operator ordering}
\label{ch:MultT}
We will now address the question of which quantum averages the path integral calculates.
To define this more clearly, consider the path-integral average \mbox{($t_1<t_2<\cdots <t_K$)}
\begin{widetext}
{\begin{multline}
\overline{\overline{\hspace{0.1ex}\alpha(t_1)\alpha(t_2)\cdots\alpha(t_K)\hspace{0.1ex}}} = \int
\frac{d^2\alpha_0 d^2\alpha_1\cdots d^2\alpha_K}{\pi ^{K+1}}
\\ \times
\alpha _K U(\alpha _K,t_K,\alpha _{K-1},t_{K-1}) 
\alpha _{K-1} U(\alpha _{K-1},t_{K-1},\alpha _{K-2},t_{K-2}) 
\cdots\alpha _1 U(\alpha _1,t_1,\alpha _{0},t_{0}) 
\rho (\alpha _0,t_0)
.
\label{eq:TWI}
\end{multline}}%
We presume that there exists a {\em rule of ordering\/} for Heisenberg operators, which we term the {\em time-symmetric\/} ordering \cite{ourpreprint} and denote ${\cal T}_W$, such that
 {\begin{align}{{
 \begin{aligned}
\overline{\overline{\hspace{0.1ex}\alpha(t_1)\alpha(t_2)\cdots\alpha(t_K) \hspace{0.1ex}}} =   \ensuremath{\big\langle
{\cal T}_W  {\hat{\mathcal A}}(t_1) {\hat{\mathcal A}}(t_2)\cdots {\hat{\mathcal A}}(t_K)
\big\rangle} .
\end{aligned}}}
\label{eq:TW}
\end{align}}%
The Heisenberg field operators are given by (\ref{eq:71b}). It is easy to obtain a recursion relation expressing
${\cal T}_W  {\hat{\mathcal A}}(t_1) {\hat{\mathcal A}}(t_2)\cdots {\hat{\mathcal A}}(t_K)$ by
${\cal T}_W  {\hat{\mathcal A}}(t_2)\cdots {\hat{\mathcal A}}(t_K)$. Comparing Eqs.\ ( \ref{eq:TWI}), (\ref{eq:TW}) to (\ref{eq:AvW}) we have
{\begin{multline}
  \ensuremath{\big[
{\cal T}_W  {\hat{\mathcal A}}_{t_0}(t_1) {\hat{\mathcal A}}_{t_0}(t_2)\cdots {\hat{\mathcal A}}_{t_0}(t_K)
\big]} (\alpha _0)
= \int
\frac{d^2\alpha_1\cdots d^2\alpha_K}{\pi ^{K}}
\\ \times
\alpha _K U(\alpha _K,t_K,\alpha _{K-1},t_{K-1}) 
\alpha _{K-1} U(\alpha _{K-1},t_{K-1},\alpha _{K-2},t_{K-2}) 
\cdots\alpha _1 U(\alpha _1,t_1,\alpha _{0},t_{0}) 
,
\label{eq:31b} 
\end{multline}}%
see also endnote \cite{EndSB}.
In this relation the dependence of the Heisenberg operators on the coincidence point is made explicit. Applying it to the product ${\cal T}_W  {\hat{\mathcal A}}_{t_1}(t_2)\cdots {\hat{\mathcal A}}_{t_1}(t_K)$ with the coincidence point set at $t_1$ we find
{\begin{multline}
  \ensuremath{\big[
{\cal T}_W  {\hat{\mathcal A}}_{t_1}(t_2)\cdots {\hat{\mathcal A}}_{t_1}(t_K)
\big]} (\alpha _1) 
= \int
\frac{d^2\alpha_2\cdots d^2\alpha_K}{\pi ^{K-1}}
\\ \times
\alpha _K U(\alpha _K,t_K,\alpha _{K-1},t_{K-1}) 
\alpha _{K-1} U(\alpha _{K-1},t_{K-1},\alpha _{K-2},t_{K-2}) 
\cdots\alpha _2 U(\alpha _2,t_2,\alpha _{1},t_{1}) 
.
\label{eq:32b} 
\end{multline}}%
Comparing Eqs.\ ( \ref{eq:31b}) and (\ref{eq:32b}) we see that
 {\begin{align}{{
 \begin{aligned}
  \ensuremath{\big[
{\cal T}_W  {\hat{\mathcal A}}_{t_0}(t_1) {\hat{\mathcal A}}_{t_0}(t_2)\cdots {\hat{\mathcal A}}_{t_0}(t_K)
\big]} (\alpha _0)
= \int
\frac{d^2\alpha_1}{\pi}\,
\alpha _1 U(\alpha _1,t_1,\alpha _{0},t_{0})
  \ensuremath{\big[
{\cal T}_W  {\hat{\mathcal A}}_{t_1}(t_2)\cdots {\hat{\mathcal A}}_{t_1}(t_K)
\big]} (\alpha _1)
.
\end{aligned}}}
\label{eq:Rec2} 
\end{align}}%
We now recall the standard phase-space correspondence,
 {\begin{align}{{
 \begin{aligned}
\alpha   \ensuremath{\big[
\hat A
\big]} (\alpha ) = \frac{1}{2}   \ensuremath{\big[
\hat a \hat A+
\hat A\hat a
\big]} (\alpha )
= \frac{1}{2}   \ensuremath{\big[
  \ensuremath{\big\{
\hat a ,\hat A
\big\}}
\big]} (\alpha )
,
\end{aligned}}}
\end{align}}%
where the curly brackets stand for the anticommutator,
$  \ensuremath{\big\{
 {\hat{\mathcal X}},  {\hat{\mathcal Y}}
\big\}} =  {\hat{\mathcal X}}  {\hat{\mathcal Y}} +  {\hat{\mathcal Y}}  {\hat{\mathcal X}} $ .
This allows us to write
 {\begin{align}{{
 \begin{aligned}
\alpha _1   \ensuremath{\big[
{\cal T}_W  {\hat{\mathcal A}}_{t_1}(t_2)\cdots {\hat{\mathcal A}}_{t_1}(t_K)
\big]} (\alpha _1)
= \frac{1}{2}   \ensuremath{\big[
  \ensuremath{\big\{
{ {\hat{\mathcal A}}_{t_1}(t_1),\cal T}_W  {\hat{\mathcal A}}_{t_1}(t_2)\cdots {\hat{\mathcal A}}_{t_1}(t_K)
\big\}}
\big]} (\alpha _1)\, ,
\end{aligned}}}
\label{eq:Rec1} 
\end{align}}%
where we have used the fact that, with the coincidence point set at $t=t_1$, the Heisenberg field operator $ {\hat{\mathcal A}}(t_1)$ coincides
with its Schr\"odinger counterpart,
 {\begin{align}{{
 \begin{aligned}
 {\hat{\mathcal A}}_{t_1}(t_1) = \hat a\, .
\end{aligned}}}
\label{eq:At1} 
\end{align}}%
Equation (\ref{eq:Rec2}) then becomes
 {\begin{align}{{
 \begin{aligned}
  \ensuremath{\big[
{\cal T}_W  {\hat{\mathcal A}}_{t_0}(t_1) {\hat{\mathcal A}}_{t_0}(t_2)\cdots {\hat{\mathcal A}}_{t_0}(t_K)
\big]} (\alpha _0)
= \frac{1}{2}\int
\frac{d^2\alpha_1}{\pi}\,
U(\alpha _1,t_1,\alpha _{0},t_{0})
  \ensuremath{\big[
  \ensuremath{\big\{
{ {\hat{\mathcal A}}_{t_1}(t_1),\cal T}_W  {\hat{\mathcal A}}_{t_1}(t_2)\cdots {\hat{\mathcal A}}_{t_1}(t_K)
\big\}}
\big]} (\alpha _1)\,.
\end{aligned}}}%
\label{eq:33b} 
\end{align}}%
We now note that Eq.\ ( \ref{eq:Bt}) is based solely on Eq.\ ( \ref{eq:95b}) and is therefore a particular case of a more general relation
 {\begin{align}{{
 \begin{aligned}
  \ensuremath{\big[
 {\hat{\mathcal U}}^{\dag} (t,t_0)
 {\hat{\mathcal X}}
 {\hat{\mathcal U}} (t,t_0)
\big]} (\alpha _0)=\int\frac{d^2\alpha}{\pi}  \ensuremath{\big[
 {\hat{\mathcal X}}
\big]} (\alpha )U(\alpha,t,\alpha_0,t_0)\, ,
\end{aligned}}}%
\label{eq:34b} 
\end{align}}%
where the operator $ {\hat{\mathcal X}}$ may be arbitrary. Applying this to (\ref{eq:33b}) we have
{\begin{multline}
  \ensuremath{\big[
{\cal T}_W  {\hat{\mathcal A}}_{t_0}(t_1) {\hat{\mathcal A}}_{t_0}(t_2)\cdots {\hat{\mathcal A}}_{t_0}(t_K)
\big]} (\alpha _0)
= \frac{1}{2}
  \ensuremath{\big[ {\hat{\mathcal U}}^{\dag} (t_1,t_0)
  \ensuremath{\big\{
{ {\hat{\mathcal A}}_{t_1}(t_1),\cal T}_W  {\hat{\mathcal A}}_{t_1}(t_2)\cdots {\hat{\mathcal A}}_{t_1}(t_K)
\big\}} {\hat{\mathcal U}} (t_1,t_0)
\big]} (\alpha _0)\\ = \frac{1}{2}
  \ensuremath{\big[
  \ensuremath{\big\{
{ {\hat{\mathcal A}}_{t_0}(t_1),\cal T}_W  {\hat{\mathcal A}}_{t_0}(t_2)\cdots {\hat{\mathcal A}}_{t_0}(t_K)
\big\}}
\big]} (\alpha _0)\,.
\label{eq:35b} 
\end{multline}}%
We have thus arrived at the desired recursion relation,
 {\begin{align}{{
 \begin{aligned}
{\cal T}_W  {\hat{\mathcal A}}(t_1) {\hat{\mathcal A}}(t_2)\cdots {\hat{\mathcal A}}(t_K)
= \frac{1}{2}
  \ensuremath{\big\{
{ {\hat{\mathcal A}}(t_1),\cal T}_W  {\hat{\mathcal A}}(t_2)\cdots {\hat{\mathcal A}}(t_K)
\big\}} ,
\end{aligned}}}
\label{eq:Rec0}
\end{align}}%
\end{widetext}%
where the dependence on the coincidence point has been dropped.

If we replace $\alpha (t_1)\to \alpha ^*(t_1)$ and
$ {\hat{\mathcal A}}(t_1)\to {\hat{\mathcal A}}^{\dag}(t_1)$ in (\ref{eq:TW}), equation (\ref{eq:Rec0}) will also hold with $ {\hat{\mathcal A}}(t_1)\to {\hat{\mathcal A}}^{\dag}(t_1)$. Furthermore, any subset of the factors $\alpha (t_2)\cdots\alpha (t_K)$ may be complex-conjugated, provided the corresponding operators under the ${\cal T}_W$-ordering are Hermitian-conjugated. As a result, we arrive at the recursive definition of the time-symmetric ordering given by Eq.\ ( \ref{eq:46b}) in section \ref{ch:TSO}. For a brief discussion of this concept we refer the reader to that section. Detailed analyses will be presented elsewhere \cite{Unpb}. 
\subsection{Commuting Heisenberg operators as a
response problem}
\label{ch:RPh}
Now we consider what happens if the quantum average we wish to calculate is not a time-symmetric one, but a time-normal one. In this paper, we only consider two-time averages ($t_0<t_1,t_2$)
 {\begin{align}{{
 \begin{aligned}
  \ensuremath{\big\langle
 {\hat{\mathcal X}}(t_1) {\hat{\mathcal Y}}(t_2)
\big\rangle} = \text{Tr}\hat \rho (t_0) {\hat{\mathcal X}}(t_1) {\hat{\mathcal Y}}(t_2) ,
\end{aligned}}}
\end{align}}%
where $ {\hat{\mathcal X}}(t), {\hat{\mathcal Y}}(t)=  {\hat{\mathcal A}}(t), {\hat{\mathcal A}}^{\dag}(t)$. A general discussion will be given elsewhere. Rather than distinguishing the cases $t_1>t_2$ and $t_1<t_2$, we assume that $t_1<t_2$ and consider two distinct averages, $  \ensuremath{\big\langle
 {\hat{\mathcal X}}(t_1) {\hat{\mathcal Y}}(t_2)
\big\rangle}$ and $  \ensuremath{\big\langle
 {\hat{\mathcal Y}}(t_2) {\hat{\mathcal X}}(t_1)
\big\rangle}$. Consider, for example, the average $  \ensuremath{\big\langle
 {\hat{\mathcal Y}}(t_2) {\hat{\mathcal A}}(t_1)
\big\rangle}$. Moving the coincidence point to $t=t_1$ and using Eqs.\ ( \ref{eq:BAv}) and (\ref{eq:At1}) we have
{\begin{multline}\hspace{0.4\columnwidth}\hspace{-98.4pt} 
  \ensuremath{\big\langle
 {\hat{\mathcal Y}}(t_2) {\hat{\mathcal A}}(t_1)
\big\rangle} = \text{Tr}  {\hat{\mathcal Y}}_{t_1}(t_2)\hat a \hat\rho (t_1) = \\
\int \frac{d^2 \alpha_1 d^2 \alpha_2}{\pi^2 }Y(\alpha_2)U(\alpha_2,t_2,\alpha_1,t_1)
  \ensuremath{\big[
\hat a \hat \rho (t_1)
\big]} (\alpha_1).
\hspace{0.4\columnwidth}\hspace{-98.4pt} 
\label{eq:61b}
\end{multline}}%
In this relation we made the dependence of $ {\hat{\mathcal Y}}(t)$ on the coincidence point explicit, cf.\ Eqs.\ ( \ref{eq:Rec2}), (\ref{eq:Rec1}) and (\ref{eq:At1}). The standard phase-space correspondences then allow us to write
 {\begin{align}{{
 \begin{aligned}
  \ensuremath{\big[
\hat a \hat \rho (t_1)
\big]} (\alpha_1) =  \ensuremath{\Big(
\alpha_1 + \frac{1}{2} \frac{\partial }{\partial \alpha_1 ^*} \Big)}
\rho (\alpha_1,t_1).
\end{aligned}}}
\end{align}}%
Using Eq.\ ( \ref{eq:rho}) to express $\rho (\alpha_1,t_1)$ we obtain
{\begin{multline}\hspace{0.4\columnwidth}\hspace{-98.4pt} 
  \ensuremath{\big\langle
 {\hat{\mathcal Y}}(t_2) {\hat{\mathcal A}}(t_1)
\big\rangle} = \int \frac{d^2 \alpha_2 d^2 \alpha_1 d^2 \alpha _0}{\pi^3 }Y(\alpha _2)
\\ \times  \ensuremath{\bigg[
 \ensuremath{\Big(
\alpha_1 - \frac{1}{2} \frac{\partial }{\partial \alpha_1 ^*} \Big)}
U(\alpha _2,t_2,\alpha _1,t_1)
\bigg]} \\ \times
U(\alpha _1,t_1,\alpha _0,t_0)\rho (\alpha _0,t_0) ,
\hspace{0.4\columnwidth}\hspace{-98.4pt} 
\label{eq:YA}
\end{multline}}%
cf.\ endnote \cite{EndSB}. Integration by parts was used to move the derivative to $U(\alpha _2,t_2,\alpha _1,t_1)$; square brackets emphasize that the differentiation does not apply to $U(\alpha _1,t_1,\alpha _0,t_0)$. Similar considerations yield
{\begin{multline}\hspace{0.4\columnwidth}\hspace{-98.4pt} 
  \ensuremath{\big\langle
 {\hat{\mathcal A}}(t_1) {\hat{\mathcal Y}}(t_2)
\big\rangle} = \int \frac{d^2 \alpha_2 d^2 \alpha_1 d^2 \alpha _0}{\pi^3 }Y(\alpha _2)
\\ \times  \ensuremath{\bigg[
 \ensuremath{\Big(
\alpha_1 + \frac{1}{2} \frac{\partial }{\partial \alpha_1 ^*} \Big)}
U(\alpha _2,t_2,\alpha _1,t_1)
\bigg]} \\ \times
U(\alpha _1,t_1,\alpha _0,t_0)\rho (\alpha _0,t_0) ,
\hspace{0.4\columnwidth}\hspace{-98.4pt} 
\label{eq:AY}
\end{multline}}%
{\begin{multline}\hspace{0.4\columnwidth}\hspace{-98.4pt} 
  \ensuremath{\big\langle
 {\hat{\mathcal A}}^{\dag}(t_1) {\hat{\mathcal Y}}(t_2)
\big\rangle} = \int \frac{d^2 \alpha_2 d^2 \alpha_1 d^2 \alpha _0}{\pi^3 }Y(\alpha _2)
\\ \times  \ensuremath{\bigg[
 \ensuremath{\Big(
\alpha_1^* - \frac{1}{2} \frac{\partial }{\partial \alpha_1 } \Big)}
U(\alpha _2,t_2,\alpha _1,t_1)
\bigg]} \\ \times
U(\alpha _1,t_1,\alpha _0,t_0)\rho (\alpha _0,t_0) ,
\hspace{0.4\columnwidth}\hspace{-98.4pt} 
\label{eq:AdY}
\end{multline}}%
{\begin{multline}\hspace{0.4\columnwidth}\hspace{-98.4pt} 
  \ensuremath{\big\langle
 {\hat{\mathcal Y}}(t_2) {\hat{\mathcal A}}^{\dag}(t_1)
\big\rangle} = \int \frac{d^2 \alpha_2 d^2 \alpha_1 d^2 \alpha _0}{\pi^3 }Y(\alpha _2)
\\ \times  \ensuremath{\bigg[
 \ensuremath{\Big(
\alpha_1^* + \frac{1}{2} \frac{\partial }{\partial \alpha_1} \Big)}
U(\alpha _2,t_2,\alpha _1,t_1)
\bigg]} \\ \times
U(\alpha _1,t_1,\alpha _0,t_0)\rho (\alpha _0,t_0) .
\hspace{0.4\columnwidth}\hspace{-98.4pt} 
\label{eq:YAd}
\end{multline}}%
We remind the reader that Eqs.\ ( \ref{eq:YA})--(\ref{eq:YAd}) hold if $t_0<t_1<t_2$. The latest operator in them is in fact arbitrary. 

Equations (\ref{eq:YA})--(\ref{eq:YAd}) are expressions of ``generalised phase-space correspondences'' discussed in section \ref{ch:TSG}. 
They are exact and not associated with the path-integral representation of the phase-space amplitude. However their most natural interpretation is in terms of path-integral averages, with the derivatives related to ``quantum jumps'' of the trajectories. 
\section{Analytical example: The Kerr oscillator}
\label{ch:ana}

As a simple illustrative example we apply the ``generalised phase-space correspondences'' (\ref{eq:wichtig}) to the Kerr oscillator introduced in section \ref{ch:ThSumm}. 
The same model was used in Ref.~\cite{ap_twa} to illustrate the effect of quantum corrections to the --W picture. This problem is exactly soluble; better still, all calculations implied by Eq.\ ( \ref{eq:wichtig}) may be completed analytically. This makes the Kerr oscillator an ideal first testing ground for our approach.

We assume that the oscillator is initially in a coherent state (this setup closely mimics the collapse-revival experiment of Ref.~\cite{bloch_collapse}):
 {\begin{align}{{
 \begin{aligned}
  \ensuremath{\big\langle
\cdots
\big\rangle} =
 {\big\langle \beta \big | 
\cdots
 \big | \beta \big\rangle}, \ \ \ \hat{a}|\beta\rangle=\beta|\beta\rangle .
\end{aligned}}}%
\label{eq:1b} 
\end{align}}%
The time-normally ordered correlation function is then easily calculated:
{\begin{multline}\hspace{0.4\columnwidth}\hspace{-98.4pt} 
G_{\textrm{H}} (t_1,t_2) = \langle \hat {\cal A}^{\dagger}(t_1)\hat {\cal A}(t_2)\rangle
\\ =|\beta|^2\exp  \ensuremath{\big\{
|\beta|^2  \ensuremath{\big[
e^{-i\kappa(t_2-t_1)}-1
\big]}
\big\}}\, .
\hspace{0.4\columnwidth}\hspace{-98.4pt} 
\label{eq:2b} 
\end{multline}}%
The subscript ``H'' distinguishes this as an exact Hilbert-space result.
This expression follows from the exact solution for the Heisenberg operators,
 {\begin{align}{{
 \begin{aligned}
\hat {\cal A}(t)=e^{-i\kappa t \hat{a}^{\dagger}\hat{a}}\hat{a},
\ \ \
\hat {\cal A}^{\dagger}(t)=\hat{a}^{\dagger}e^{i\kappa t \hat{a}^{\dagger}\hat{a}},
\end{aligned}}}%
\label{eq:Heiop} 
\end{align}}%
so that
\begin{align}
&&G_{\textrm{H}}(t_1,t_2)=\langle\beta|\hat{a}^{\dagger}e^{it_1\kappa\hat{a}^{\dagger}\hat{a}}e^{-it_2\kappa\hat{a}^{\dagger}\hat{a}}\hat{a}|\beta\rangle\, .
\end{align}
Equation (\ref{eq:2b}) is found by expanding the exponents in power series and recalling the expansion of a coherent state over the number states.

Calculations associated with Eq.\ ( \ref{eq:wichtig}) take more effort but are also quite straightforward.
The equation for phase-space trajectories
in the truncated Wigner representation is given by (\ref{eq:5b}) with $s(t)=0$. 
Phase-space evolution only affects the phase of $\alpha (t)$, so that $|\alpha(t)|^2=|\alpha(0)|^2$. With this observation Eq.\ ( \ref{eq:5b}) is solved trivially,
 {\begin{align}{{
 \begin{aligned}
\alpha(t)=\alpha(0)\exp\big\{-i\kappa t\big(|\alpha(0)|^2-1\big)\big\}.
\end{aligned}}}%
\label{eq:6b} 
\end{align}}%
Stochasticity only enters the picture through the initial condition for $\alpha(t)$, distributed with probability
 {\begin{align}{{
 \begin{aligned}
W(\alpha,\alpha^*)=\frac{2}{\pi}\exp\{-2|\alpha(0)-\beta|^2\}.
\end{aligned}}}%
\label{eq:7b} 
\end{align}}%
Strictly speaking, $W(\alpha,\alpha^*)$ is the {\em Weyl-ordered quasiprobability distribution\/}, or {\em Wigner function\/}, of the state $  \ensuremath{ |
\beta
 \rangle } $, but with $W(\alpha,\alpha^*)\geq 0$ such formal niceties may be disregarded.

By making use of Eqs.\ ( \ref{eq:6b}) and (\ref{eq:7b}) we find for the symmetrically-ordered correlation function: 
\begin{widetext}
 {\begin{gather}{{
 \begin{gathered}
G_{\textrm{W}}(t_1,t_2) = \overline{\alpha^*(t_1)\alpha(t_2)}
 = \frac{|\beta|^2+\frac{1}{2}-\frac{i\kappa}{4}(t_1-t_2)}
{  \ensuremath{\big[
1-\frac{i\kappa}{2}(t_1-t_2)
\big]} ^3} 
\exp\bigg[i\kappa(t_1-t_2)\frac{|\beta|^2-1+\frac{i\kappa}{2}(t_1-t_2)}{1-\frac{i\kappa}{2}(t_1-t_2)}\bigg].
\end{gathered}}}%
\label{eq:8b} 
\end{gather}}%
For $t_1=t_2=0$ we have $G_{\textrm{W}}(0,0)=|\beta|^2+1/2$ as expected.
The response terms in Eq.\ ( \ref{eq:wichtig}) are also easily calculated, leading to, 
 {\begin{align}{{
 \begin{aligned}
\overline{\hspace{0.1ex}\frac{\partial{ \alpha(t_2)}}{\partial \alpha(t_1)}\hspace{0.1ex}}=
\frac{1+\frac{i\kappa}{2}(t_1-t_2)(2|\beta|^2-1)}
{  \ensuremath{\big[
1-\frac{i\kappa}{2}(t_1-t_2)
\big]}^3} \exp\bigg[i\kappa(t_1-t_2)\frac{|\beta|^2-1+\frac{i\kappa}{2}(t_1-t_2)}
{1-\frac{i\kappa}{2}(t_1-t_2)}\bigg] , \ \ t_2>t_1.
\end{aligned}}}%
\label{eq:9b} 
\end{align}}%
The quantity $\Big[\overline{\hspace{0.1ex}{\partial{ \alpha(t_1)}}/{\partial \alpha(t_2)}\hspace{0.1ex}}\Big]^*$ for $t_1>t_2$ is given by the same expression. 
Unlike Eq.\ ( \ref{eq:2b}) which is exact, Eqs.\ ( \ref{eq:8b}) and (\ref{eq:9b}) are approximations within the truncated Wigner approach. Combining them, we find the truncated-Wigner approximation to the normally ordered correlation function,
\begin{align}
G_{\textrm{N}}(t_1,t_2) =
\frac{|\beta|^2}{  \ensuremath{\big[
1-\frac{i\kappa}{2}(t_1-t_2)
\big]}^2} \exp\bigg[i\kappa(t_1-t_2)\frac{|\beta|^2-1+\frac{i\kappa}{2}(t_1-t_2)}
{1-\frac{i\kappa}{2}(t_1-t_2)}\bigg] \, .
\label{eq:10b} 
\end{align}
Both the exact formula and the approximate formula for the time-normally ordered correlation function depend on the time difference $\Delta t = t_1-t_2$.

\begin{figure*}
\includegraphics[angle=270,width=0.99\textwidth]{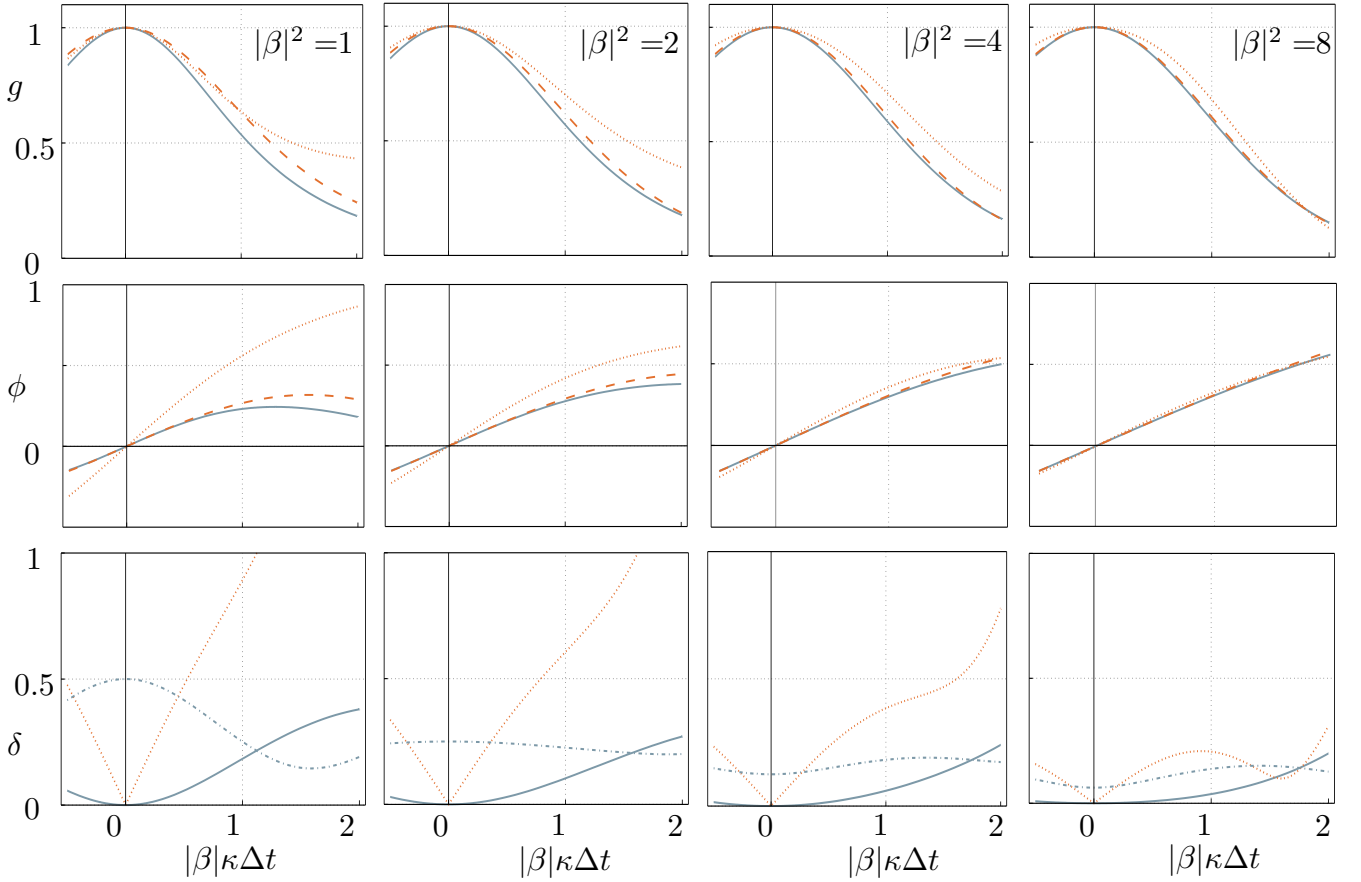}
\caption{Normally-ordered correlation function for the Kerr-oscillator. Top and middle rows: Comparison between the exact solution $G_{\textrm{H}} (t_1,t_2)$ (dashed line), the truncated-Wigner solution $G_{\textrm{N}}(t_1,t_2)$ (solid line) and the ``naively corrected'' solution $G_{\textrm{C}}(t_1,t_2)$ (dotted line) for $|\beta|^2=1,~2,~4,~8$ (from left to right). Top row of graphs depicts the scaled moduli and the middle row --- scaled phases, cf.\ Eqs.\ ( \ref{eq:39b}), (\ref{eq:40b}). Bottom row: relative erros of various approximations to the normally-ordered correlation function (\ref{eq:41b}). Truncated-Wigner approximation $G_{\textrm{N}}(t_1,t_2)$ (solid lines), uncorrected (symmetric) truncated-Wigner solution $G_{\textrm{W}}(t_1,t_2)$ (dash-dotted lines), and the ``naively corrected'' solution $G_{\textrm{C}}(t_1,t_2)$ (dotted lines). All quantities are plotted against the scaled time difference $|\beta|\kappa \Delta t$.}
\label{fig:Ana}
\end{figure*}

We can demonstrate the accuracy of this approximate result by
considering the series expansion,
 {\begin{align}{{
 \begin{aligned}
\log\big[
G_{\text{N}}(\Delta t)/G_{\text{H}}(\Delta t)
\big] =-\frac{\kappa^2 \Delta t^2}{4} -\frac{i}{12} \left(|\beta|^2+1\right) \kappa^3 \Delta t^3+\frac{1}{96} \left(8 |\beta|^2+3\right) \kappa^4 \Delta t^4 +O\left(\Delta t^5\right),
\end{aligned}}}%
\label{eq:13b} 
\end{align}}%
\end{widetext}%
from which we see that Eq.\ (\ref{eq:10b}) is a good approximation if $\kappa |\Delta t| \ll 1$, $|\beta|\kappa |\Delta t| \sim 1$. In other words, it holds over the collapse time scale, $|\Delta t| \sim 1/|\beta|\kappa$, but fails over the revival time scale,
$|\Delta t| \sim 1/\kappa$.

In Fig.\ \ref{fig:Ana} we compare the exact function $G_{\textrm{H}} (t_1,t_2)$ to its truncated-Wigner approximation $G_{\textrm{N}}(t_1,t_2)$ and to the ``naively'' corrected correlation function
 {\begin{align}{{
 \begin{aligned}
G_{\textrm{C}}(t_1,t_2)=G_{\textrm{W}}(t_1,t_2)-1/2 .
\end{aligned}}}%
\label{eq:38b} 
\end{align}}%
(the latter corresponds to using the free-field commutator
in place of the Heisenberg one) for different values of $\beta $.
Plots in the top and middle rows depict, respectively, the scaled modulus
 {\begin{align}{{
 \begin{aligned}
g(t_1,t_2)=|G(t_1,t_2)|/|\beta|^2
\end{aligned}}}%
\label{eq:39b} 
\end{align}}%
and the scaled phase
 {\begin{align}{{
 \begin{aligned}
\phi(t_1,t_2)=\mathrm{arg}\,G(t_1,t_2)/|\beta|\pi
\end{aligned}}}%
\label{eq:40b} 
\end{align}}%
of the correlation function for $G(t_1,t_2) = G_{\textrm{H}}(t_1,t_2), G_{\textrm{N}}(t_1,t_2),G_{\textrm{C}}(t_1,t_2)$.
Plots in the bottom row show the relative error,
\begin{align}
\delta(t_1,t_2)=\bigg|\frac{G(t_1,t_2)}{G_{\textrm{H}}(t_1,t_2)}-1\bigg| .
\label{eq:41b} 
\end{align}
Here we also include the symmetric correlation function $G(t_1,t_2) = G_{\textrm{W}}(t_1,t_2)$.
Each column of plots corresponds to one value of $\beta $: from left to right, $|\beta |^2 = 1,2,4,$ and 8.
All graphs are plotted against the scaled time difference $|\beta|\kappa\Delta t$.

We see that the accuracy of Eq.\ (\ref{eq:10b}) is always superior to that of the uncorrected as well as the naively corrected
symmetric average.
The response correction brings the truncated Wigner prediction into excellent agreement with the true solution for $|\beta|^2\geq 2$, but is not very accurate for $|\beta|^2=1$. The lack of accuracy for $|\beta|^2=1$ is not an unexpected result as the truncation process is generally thought to be justifiable as long as the number of quanta is significantly greater than the number of modes~\cite{Norrie}. What is perhaps surprising here is how accurate the approximation becomes for $|\beta |^2$ as small as $2$, although we must remark that accuracy in calculating one particular operator moment does not imply accuracy in the calculation of all possible moments. It is worthy of reminding the reader that inaccuracies in Eqs.\ ( \ref{eq:8b})--(\ref{eq:10b}) are solely due to the approximate nature of the truncated Wigner approach. By itself, Eq.\ ( \ref{eq:wichtig}) is exact.

\section{Numerical example: The Bose-Hubbard chain}
\label{ch:num}

In this section we apply our method to the one-dimensional Bose-Hubbard model. This model describes, in particular, neutral bosons in deep optical lattices and can be readily realized in experiments (see Ref.~\cite{bloch_review} for an overview). The Bose-Hubbard model is described by the Hamiltonian,
{\begin{multline}\hspace{0.4\columnwidth}\hspace{-98.4pt} 
\hat{H} =\hbar \sum_{k=1}^{N}\bigg[\omega_0\hat{n}_k+\frac{\kappa}{2}\hat{n}_k
\big(\hat{n}_k-1\big)
\\
-J\big(\hat{a}^{\dagger}_k\hat{a}_{k+1}+\hat{a}^{\dagger}_{k+1}\hat{a}_k
\big)\bigg]
\hspace{0.4\columnwidth}\hspace{-98.4pt} 
\label{eq:Hamilton} 
\end{multline}}%
with $\hat{n}_k=\hat{a}_k^{\dagger}\hat{a}_k$ and $\hat{a}_k$, $\hat{a}^{\dagger}_k$ being the standard creation-annihilation pair for the $k$-th site,
\begin{align}
\big[\hat{a}_k,\hat{a}_{k'}^{\dagger}\big]=\delta_{kk'} ,
\end{align}
with $\delta_{kk'}$ being the Kronecker delta. For simplicity we will consider a closed ring with periodic boundary conditions. The indices are to be understood modulo $N$, $\hat{a}_{N+1}=\hat{a}_1$. 
All definitions given in section \ref{ch:TSH} for the Kerr oscillator apply with the replacements 
$\hat a\to \hat a_k$, 
$\hat a^{\dag}\to \hat a_k^{\dag}$, 
$ {\hat{\mathcal A}}(t)\to  {\hat{\mathcal A}}_k(t)$, and 
$ {\hat{\mathcal A}}^{\dag}(t)\to  {\hat{\mathcal A}}_k^{\dag}(t)$. 
The truncated-Wigner equations of motion for the system described by the Hamiltonian (\ref{eq:Hamilton}) 
read 
 {\begin{align}{{
 \begin{aligned}
i\dot{\alpha_k} =\kappa\big(\big|\alpha_k\big|^2
-1\big)\alpha_k-J\big(\alpha_{k+1}+\alpha_{k-1}\big)\, .
\end{aligned}}}
 \label{eq:LangevinX} 
\end{align}}%
These equations follow from the traditional phase-space methods \cite{Wminus} and may be generalised to multitime averages using the path-integral 
approach in section \ref{ch:Theory}. 

Our aim is now to calculate the two-time normally ordered correlation function,
 {\begin{align}{{
 \begin{aligned}
G_{\textrm{H}kk'}(t_1,t_2)=   \ensuremath{\big\langle
\hat {\cal A}_k^{\dagger}(t_1)\hat {\cal A}_{k'}(t_2)
\big\rangle} .
\end{aligned}}}%
\label{eq:14b} 
\end{align}}%
For two or more sites, this is a real problem: the problem is not exactly soluble, nor can the calculations in the Wigner approach be done analytically. For the latter, a natural choice is numerics in phase-space; after all, making real systems amenable to such methods is our ultimate goal. We employ the obvious generalisation of the one-mode formula (\ref{eq:wichtig}):
\begin{widetext}
\begin{align}
G_{\textrm{H}kk'}(t_1,t_2)\approx G_{\textrm{N}kk'}(t_1,t_2) = \left\{
\begin{array}{ll}
\overline{\alpha^*_k(t_1)\alpha_{k'}(t_2)-
\displaystyle\frac{1}{2} \frac{\partial {\alpha_{k'}(t_2)}}{\partial \alpha_{k}(t_1)}}\,,
&t_1<t_2,\\ \ & \ \\
\displaystyle\overline{\alpha^*_k(t_1)\alpha_{k'}(t_2)-\frac{1}{2} \bigg[\frac{\partial {\alpha_{k}(t_1)}}{\partial \alpha_{k'}(t_2)}\bigg]^*}\,,&t_1>t_2\, .
\end{array}
\right .
\label{eq:wichtigk}
\end{align}
\end{widetext}%
We write this formula as an approximate one implying that the numerics are done with the truncated Wigner representation. In this case the averaging on the RHS reduces to that over the initial Wigner function, while the trajectories obey Eqs.\ ( \ref{eq:LangevinX}). Were the bar replaced by the double bar denoting a full path-integral quasiaverage, Eq.\ ( \ref{eq:wichtigk}) would become exact. 

Importantly, implementing Eq.\ ( \ref{eq:wichtigk}) does not require independent ``quantum jumps'' at every time step. In fact a jump at zero time suffices. For each trajectory one can then use the chain formula
{\begin{multline}\hspace{0.4\columnwidth}\hspace{-98.4pt} 
\frac{\partial \alpha _k(t_2)}{\partial \alpha_{k'}(t_1)} =
\sum_{k''}   \ensuremath{\Bigg[
\frac{\partial \alpha _k(t_2)}{\partial \alpha_{k''}(t_0)}
\frac{\partial \alpha_{k''}(t_0)}{\partial \alpha _{k'}(t_1)}
\\ +
\frac{\partial \alpha _k(t_2)}{\partial \alpha^*_{k''}(t_0)}
\frac{\partial \alpha^*_{k''}(t_0)}{\partial \alpha _{k'}(t_1)}
\Bigg]}
\hspace{0.4\columnwidth}\hspace{-98.4pt} 
\label{eq:36b} 
\end{multline}}%
with $t_0=0$.
The quantities ${\partial \alpha_{k''}(t_0)}/{\partial \alpha _{k'}(t_1)}$ and ${\partial \alpha^*_{k''}(t_0)}/{\partial \alpha _{k'}(t_1)}$ are found by inverting the matrix comprising
${\partial \alpha _k(t_1)}/{\partial \alpha_{k'}(t_0)}$,
${\partial \alpha^* _k(t_1)}/{\partial \alpha_{k'}(t_0)}$,
${\partial \alpha _k(t_1)}/{\partial \alpha^*_{k'}(t_0)}$, and
${\partial \alpha^* _k(t_1)}/{\partial \alpha^*_{k'}(t_0)}$.
Further details can be worked out by complementing Eq.\ ( \ref{eq:36b}) with similar chain relations for ${\partial \alpha^* _k(t_2)}/{\partial \alpha_{k'}(t_1)}$,
${\partial \alpha _k(t_2)}/{\partial \alpha^*_{k'}(t_1)}$ and
${\partial \alpha^* _k(t_2)}/{\partial \alpha^*_{k'}(t_1)}$, and using
 {\begin{align}{{
 \begin{aligned}
&\frac{\partial \alpha _k(t_1)}{\partial \alpha_{k'}(t_1)} =
\frac{\partial \alpha^* _k(t_1)}{\partial \alpha^*_{k'}(t_1)}
= \delta _{kk'} , \\
&\frac{\partial \alpha^* _k(t_1)}{\partial \alpha_{k'}(t_1)} =
\frac{\partial \alpha _k(t_1)}{\partial \alpha^*_{k'}(t_1)} = 0 .
\end{aligned}}}%
\label{eq:37b} 
\end{align}}%
Numerical implementation of Eq.\ ( \ref{eq:wichtigk}) thus requires a minimum of $2N+1$ trajectories run in parallel, for every initial condition generated from the distribution
 {\begin{align}{{
 \begin{aligned}
W({\mbox{\rm\boldmath$\alpha$}}(0),{\mbox{\rm\boldmath$\alpha$}}^*(0))= \ensuremath{\bigg(
\frac{2}{\pi}
 \bigg)}^N
\prod _{k=1}^N\exp  \ensuremath{\big[
-2|\alpha_k(0)-\beta_k|^2
\big]}.
\end{aligned}}}%
\label{eq:22b} 
\end{align}}%
This formula implies that the initial condition (the Heisenberg $\rho $-matrix) we use when evaluating (\ref{eq:wichtigk}) is a direct product of coherent states,
 {\begin{align}{{
 \begin{aligned}
  \ensuremath{\big| {\mbox{\rm\boldmath$\beta $}}
\big\rangle }
=
  \ensuremath{\big| \beta_1 \big\rangle } \otimes
  \ensuremath{\big| \beta_2 \big\rangle } \otimes \cdots \otimes
  \ensuremath{\big| \beta_N \big\rangle } .
\end{aligned}}}%
\label{eq:23b} 
\end{align}}%
For better numerical performance we implemented four independent shifts per mode, requiring $4N+1$ trajectories per ``coin toss.'' Such numerical cost is obviously not prohibitive. 

\begin{figure*}
\begin{center}
\includegraphics[width=0.46\textwidth]{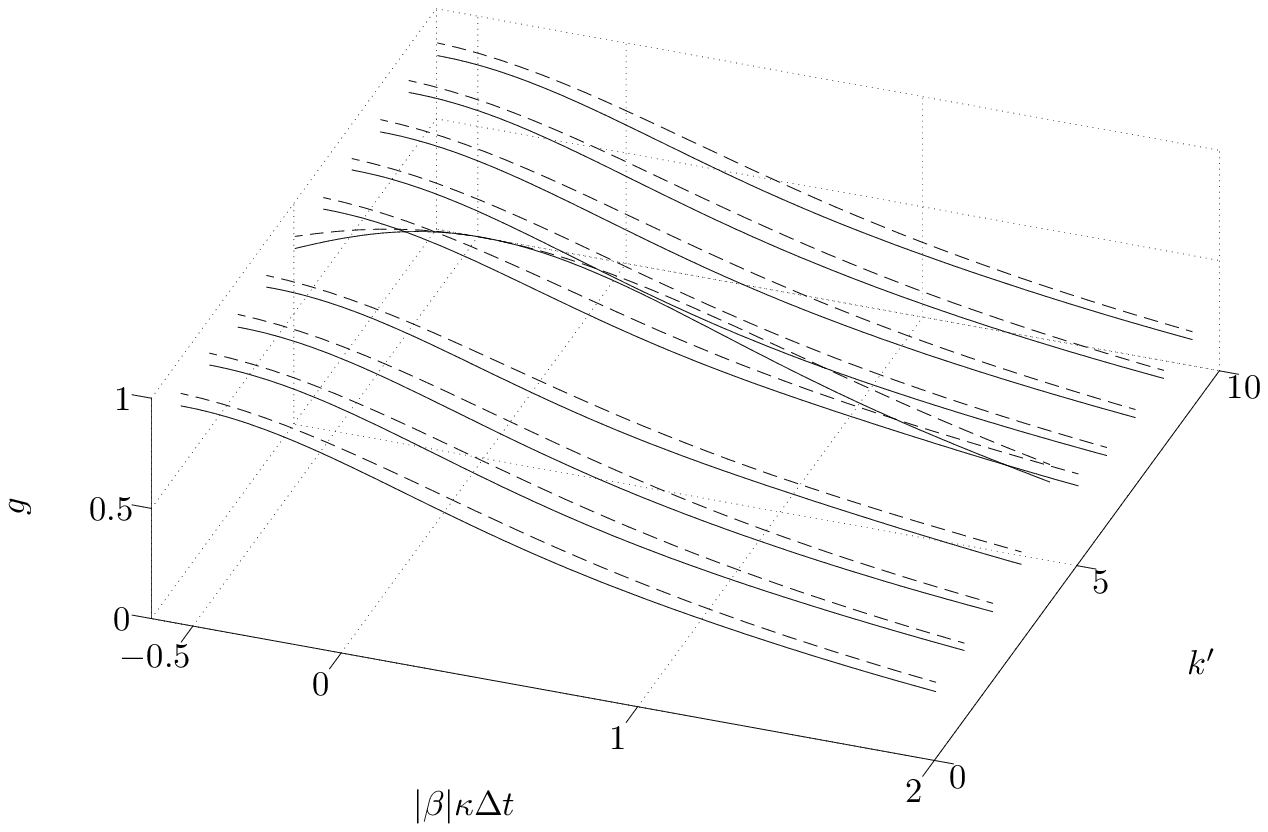}
\hspace{0.01\textwidth}
\includegraphics[width=0.46\textwidth]{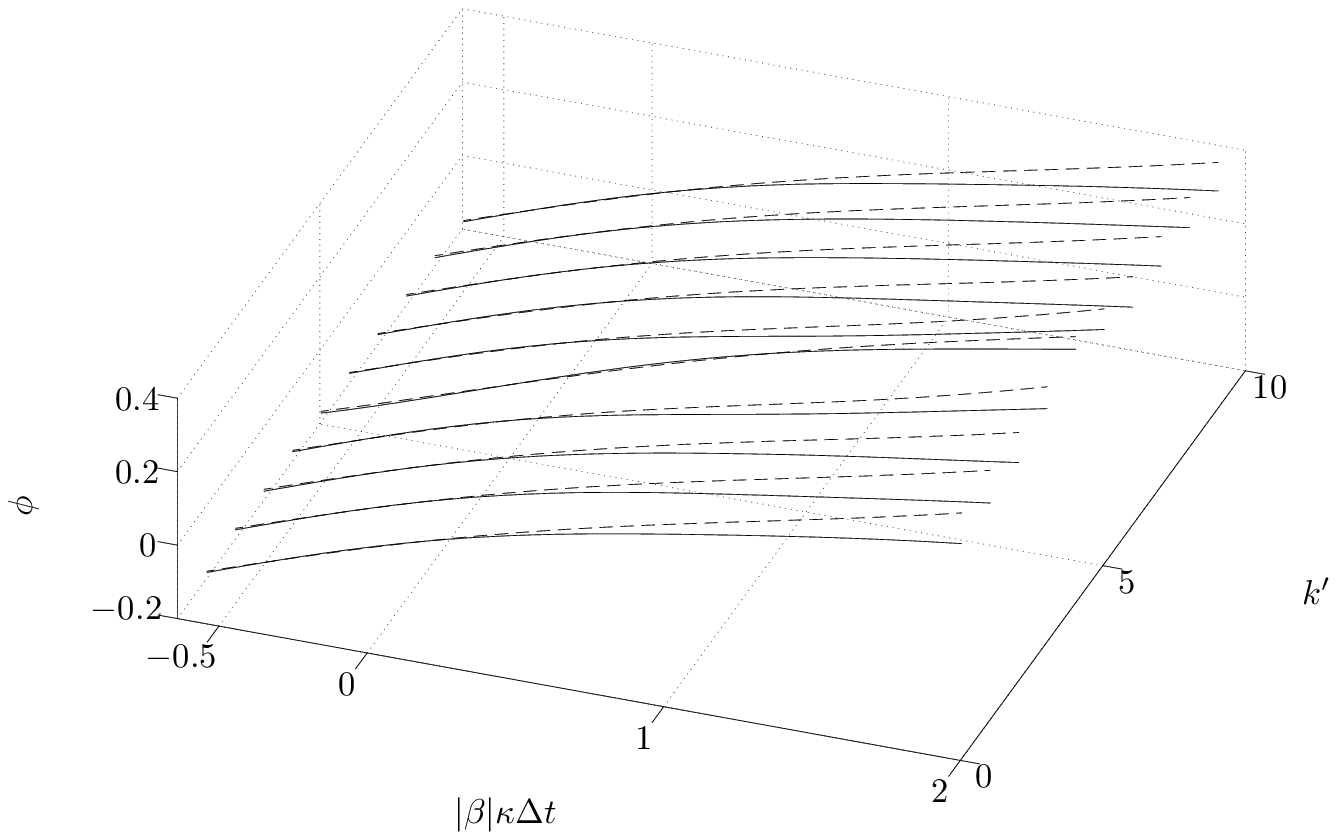}
\caption{%
The normally-ordered correlation function for the Bose-Hubbard chain of 10 sites simulated using the truncated-Wigner approximation with response correction (solid line) and the TEBD-method (dashed line) for comparision. Left: the scaled modulus $g_{\textrm{N}kk'}(t_1,t_2)$, right: the scaled phase $\phi_{\textrm{N}kk'}(t_1,t_2)$, cf.\ Eqs.\ ( \ref{eq:18b}), (\ref{eq:19b}), for $k=5$ and $k'=1,\cdots,10$. The initial condition $\beta _k=\sqrt{2}, k=1,\cdots,10$; $J=0.1$ and $\kappa=1$. Graphs are plotted versus the scaled time $|\beta|\kappa \Delta t$ with $t_2$ chosen arbitrarily as $t_2=0.45$.%
}
\label{fig:10D}
\end{center}
\end{figure*}%
As we wish to have a ``reference point'' against which to compare our results, we can use either exact diagonalization, which forces us to limit the number of sites in the Bose-Hubbard chain to two or three or to use the time-evolving block decimation algorithm (TEBD)~\cite{TEBD}. The latter assumes small entanglement in the chain which is justified for not too large times and sufficiently small systems. In Fig.\,\ref{fig:10D} we plot the result of the truncated-Wigner calculation of the normally-ordered correlation function for $N_s=10$ sites and compare this to TEBD simulations. As the TEBD algorithm favors open boundary conditions we here (and only here) use these conditions. The figure shows the scaled modulus,
 {\begin{align}{{
 \begin{aligned}
g_{\textrm{N}kk'}(t_1,t_2)=\frac{|G_{\textrm{N}kk'}(t_1,t_2)|}{|\beta|^2},
\end{aligned}}}%
\label{eq:18b} 
\end{align}}%
(left), and the scaled phase,
 {\begin{align}{{
 \begin{aligned}
\phi_{\textrm{N}kk'}(t_1,t_2)=\frac{\mathrm{arg}\,G_{\textrm{N}kk'}(t_1,t_2)}{|\beta|\pi},
\end{aligned}}}%
\label{eq:19b} 
\end{align}}%
(right), of the correlation function for $k=5$ and $k'$ ranging from 1 to 10.
In other words, each line in Fig.\ \ref{fig:10D} corresponds to correlations between site 5 and either itself or some other site.
All quantities are plotted versus the scaled time difference $|\beta|\kappa \Delta t$.
The inital condition was chosen as the same coherent state $  \ensuremath{ |
\beta _k
 \rangle } $ in all modes with $\beta_k=\sqrt{2}, \ k=1,\cdots,10$ (i.e., two quanta per mode). The hopping strength was set to $J=0.1$ and the interaction strength to $\kappa=1$.
The time $t_2$ was chosen arbitrarily as $t_2=0.45$.
The average for the truncated-Wigner method was over $80,000$ runs. 

One recognizes a rather good agreement. Since the TEBD calculations are expensive we will resort in the following examples to the case of two and three modes, where direct numerical calculations in the full Hilbert space remain doable.
\begin{figure*}
\begin{center}
\includegraphics[width=0.46\textwidth]{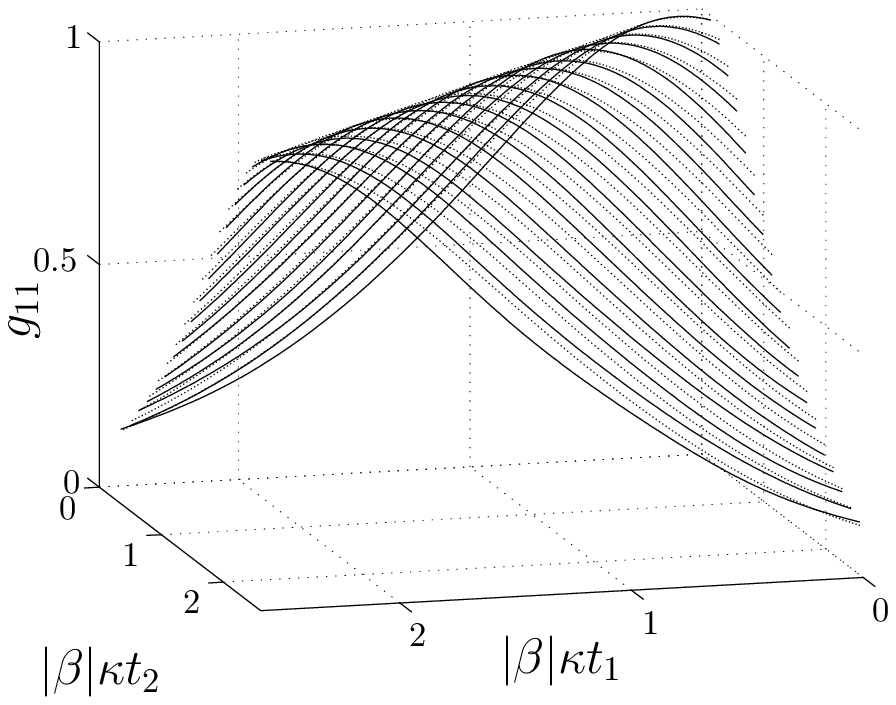}
\hspace{0.01\textwidth}
\includegraphics[width=0.46\textwidth]{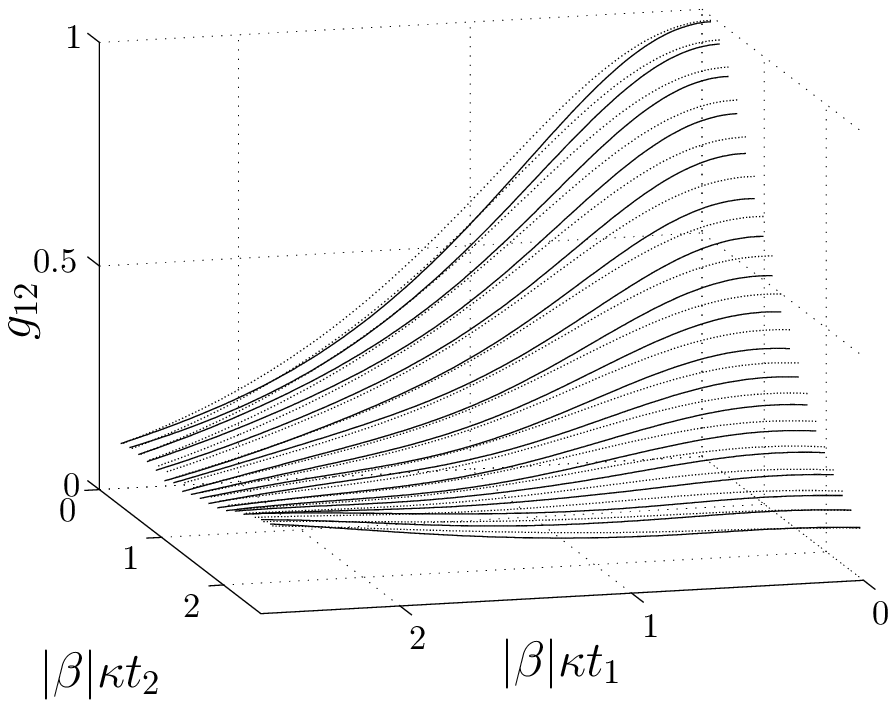}\\
\vspace*{.5cm}
\includegraphics[width=0.46\textwidth]{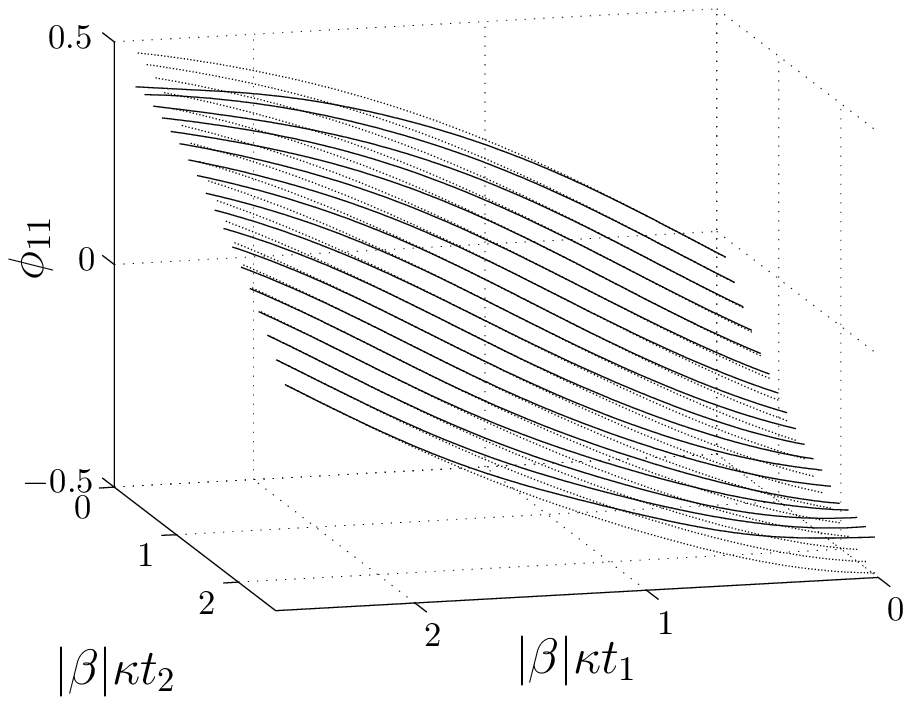}
\hspace{0.01\textwidth}
\includegraphics[width=0.46\textwidth]{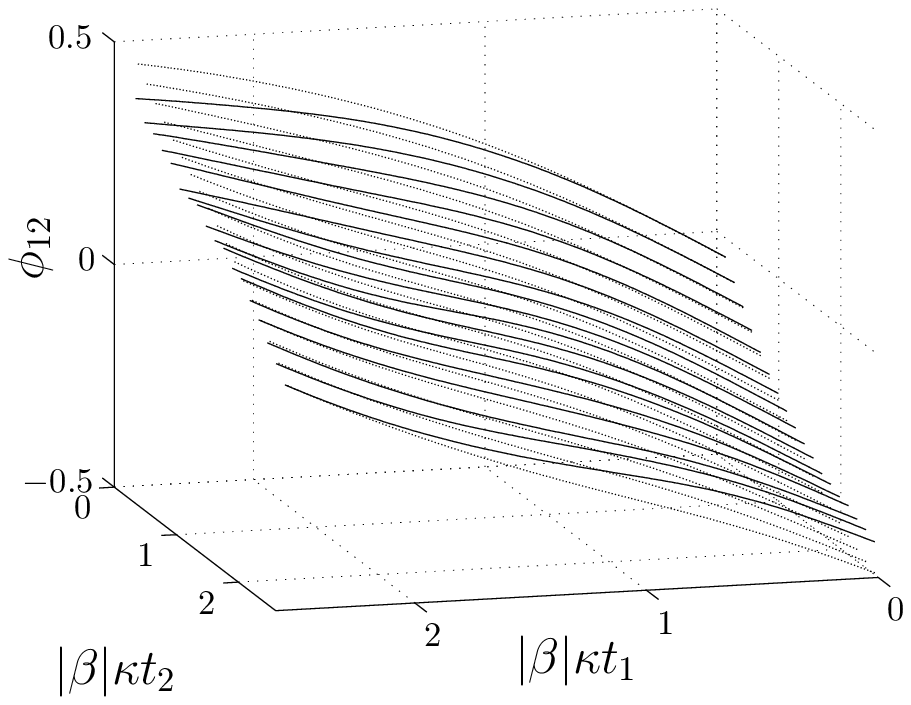}
\caption{%
Comparison of the truncated-Wigner results (solid lines) to Hilbert-space ones (dashed lines) for the normally-ordered correlation function of the Bose-Hubbard chain of two sites. Top row: the scaled moduli $g_{11}(t_1,t_2)$ (left) and $g_{12}(t_1,t_2)$ (right), bottom row: the scaled phases $\phi_{11}(t_1,t_2)$ (left) and $\phi_{12}(t_1,t_2)$ (right).
The initial condition is $\beta _1=\beta _2=\sqrt{2}$; $J=0.1$ and $\kappa=1$. The time $t_1$ changes continuously while $t_2$ is limited to discrete values ranging from 0.1 to 1.8.
}
\label{fig:2D}
\end{center}
\end{figure*}

In Fig.\ \ref{fig:2D} we compare the results of the phase-space simulations (solid lines) to those in the Hilbert space (dashed lines) for the two mode case. The plotted quantities are
$g_{11}(t_1,t_2)$ (top left),
$g_{12}(t_1,t_2)$ (top right),
$\phi_{11}(t_1,t_2)$ (bottom left), and
$\phi_{12}(t_1,t_2)$ (bottom right).
That is, the top row shows the modulus while the bottom row---the phase of the normally-ordered correlator; the left column depicts the same-site, while the right column---the neighbour-to-neighbour correlations.
Dependence of all quantities on $t_1$ and $t_2$ is expressed naturally by 3D plots. However, while $t_1$ changes continuously, $t_2$ is limited to discrete values, $t_2=0.1,\,0.2,\,\ldots,\,1.8$. The initial condition is a coherent state $  \ensuremath{ |
\beta \rangle } $ with $\beta =\sqrt{2}$ in each mode, the hopping strength is set to $J=0.1$ and the interaction strength to $\kappa=1$. The average is over 80,000 runs. We see that the conditions imposed by number conservation,
$g_{11}(t,t)=1$, $\phi_{11}(t,t)=0$, are clearly met in Fig.\ \ref{fig:2D}. Furthermore, the agreement between Hilbert space and the truncated-Wigner approximation is very good. The modulus of the correlation functions is always well reproduced. The phase appears to be not as well reproduced, but this impression is deceptive. In fact, error in phase increases when the modulus becomes small. Even in this case, the truncated-Wigner approach gives a reasonably good approximation to the phase of the correlation functions.

\begin{turnpage}
\begin{figure*}
\includegraphics[height=13cm]{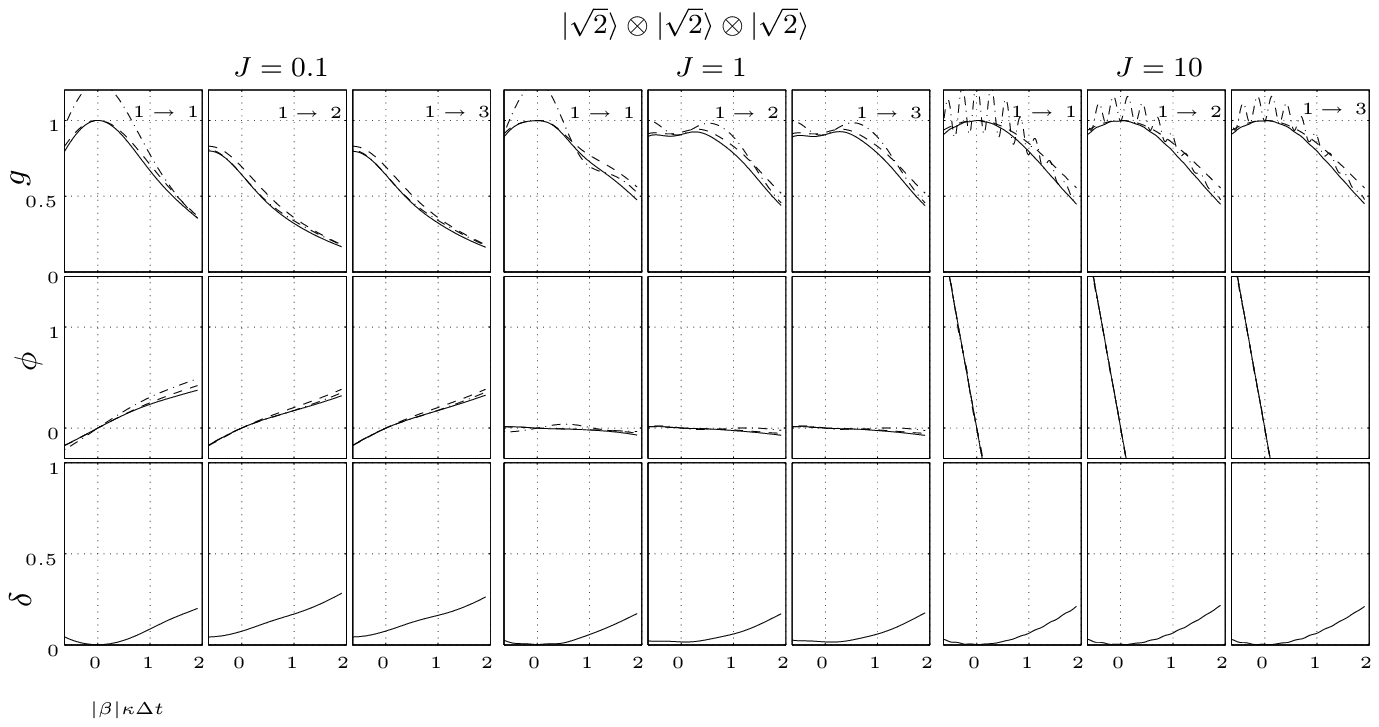}
\caption{The normally-ordered correlation function for the three mode Bose-Hubbard chain for the initial condition (\ref{eq:20b}), $\kappa =1$ and $J=0.1,1,10$. Top row: the quantities $g_{11}(t_1,t_2)$, $g_{12}(t_1,t_2)$ and $g_{13}(t_1,t_2)$; grouping of graphs with respect to the values of $J$ is self-explanatory. Truncated-Wigner calculations (solid lines), Hilbert-space results (dashed lines), and the uncorrected (symmetric) correlation function (dash-dotted lines). Middle row: the corresponding scaled phases $\phi_{11}(t_1,t_2)$, $\phi_{12}(t_1,t_2)$ and $\phi_{13}(t_1,t_2)$. Bottom row: the relative error of the truncated-Wigner simulation compared to the Hilbert-space result. All quantities are plotted versus the scaled time difference $|\beta|\kappa \Delta t$ with $t_2$ choosen arbitrarily as $t_2=0.45$.}
\label{fig:GleicheBeta}
\end{figure*}
\end{turnpage}
\begin{turnpage}
\begin{figure*}
\includegraphics[height=13cm]{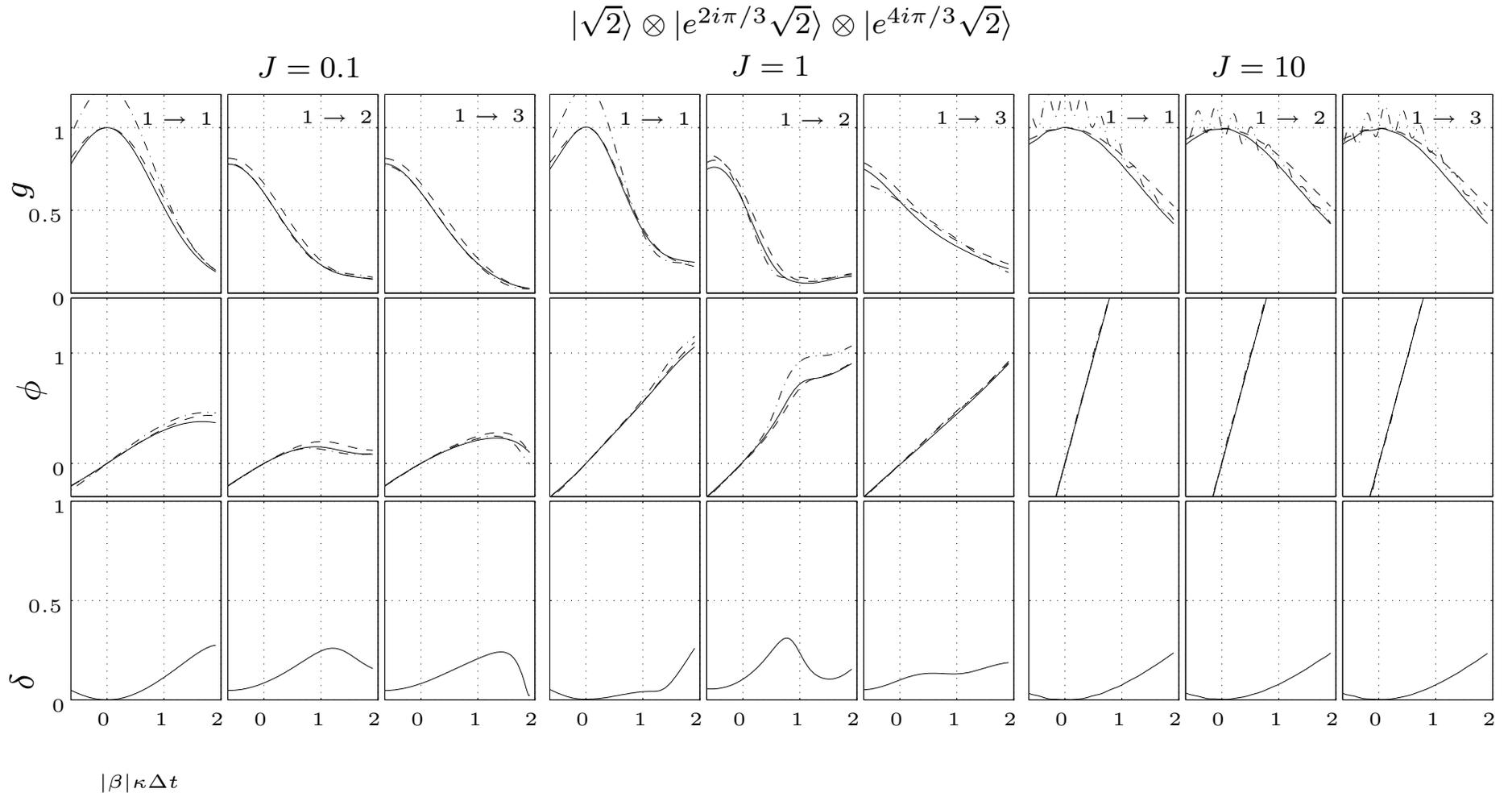}
\caption{The same as in Fig.\ \ref{fig:GleicheBeta} but for the initial condition (\ref{eq:21b}).}
\label{fig:VerschiedeneBeta}
\end{figure*}
\end{turnpage}
A detailed comparison between the truncated-Wigner and Hilbert-space calculations for three sites may be seen in Figs.\ \ref{fig:GleicheBeta} and \ref{fig:VerschiedeneBeta}. The initial condition in Fig.\ \ref{fig:GleicheBeta} was chosen as three identical coherent states,
 {\begin{align}{{
 \begin{aligned}
  \ensuremath{\big|
{\mbox{\rm\boldmath$\beta $}}
\big\rangle }
=
  \ensuremath{\big| \beta \big\rangle } \otimes
  \ensuremath{\big| \beta \big\rangle } \otimes
  \ensuremath{\big| \beta \big\rangle }, \ \ \ \textrm{(Fig.\ \ref{fig:GleicheBeta})}
\end{aligned}}}%
\label{eq:20b} 
\end{align}}%
with $\beta =\sqrt{2}$, while in Fig.\ \ref{fig:VerschiedeneBeta} the coherent states differ in phases,
 {\begin{align}{{
 \begin{aligned}
  \ensuremath{\big|
{\mbox{\rm\boldmath$\beta $}}
\big\rangle }
=
  \ensuremath{\big| \beta \big\rangle } \otimes
  \ensuremath{\big| \beta e^{2i\pi/3}\big\rangle } \otimes
  \ensuremath{\big|\beta e^{4i\pi/3}\big\rangle }, \ \ \ \textrm{(Fig.\ \ref{fig:VerschiedeneBeta})}
\end{aligned}}}%
\label{eq:21b} 
\end{align}}%
with the same $\beta $. Calculations were performed for $\kappa =1$ and $J=0.1,1,10$. In both figures, the top row of graphs shows the quantities $g_{11}(t_1,t_2)$, $g_{12}(t_1,t_2)$ and $g_{13}(t_1,t_2)$ plotted versus the scaled
time difference $|\beta|\kappa (t_1-t_2)$, with $t_2$ chosen arbitrarily as $t_2=0.45$. The grouping of graphs in Figs.\ \ref{fig:GleicheBeta} and \ref{fig:VerschiedeneBeta} with respect to the values of $J$ is self-explanatory. The results of truncated-Wigner calculations using Eq.\ ( \ref{eq:wichtigk}) are shown as solid lines, the dashed lines represent the Hilbert-space results, and the dash-dotted lines---the uncorrected (symmetrically-ordered) correlation function. The middle row of graphs depicts the corresponding scaled phases $\phi_{11}(t_1,t_2)$, $\phi_{12}(t_1,t_2)$ and $\phi_{13}(t_1,t_2)$. The bottom row represents the relative error of the truncated-Wigner simulation compared to the Hilbert-space result,
\begin{align}
\delta_{kk'}(t_1,t_2)=\bigg|\frac{G_{\textrm{N}kk'}(t_1,t_2)}{G_{\textrm{H}kk'}(t_1,t_2)}-1\bigg| .
\end{align}
Phase space averages were taken over $80,000$ runs.

Once again, we see that the requirements imposed by number conservation are met in all results. For $t_1=t_2$, $g_{11}(t_1,t_2)=1$ and $\phi_{11}(t_1,t_2)=0$.

The importance and accuracy of the response correction manifests itself pretty impressively for $J=10$ where the uncorrected phase-space solutions oscillate. This is not a numerical artifact because oscillations occur irrespective of the time step of the numerical integration. These oscillations cancel out with similar ocillations in the response term leaving the correlation function smooth and in good agreement with the Hilbert-space result.

Looking at the error plots one can recognize that the method performs well over the collapse time scale. It is worthy of stressing that we look at the relative and not at the absolute error. The error tends to get large for $\Delta t \rightarrow 1/|\beta|\kappa$ but actually this is only due to the fact that the absolute value of the correlation function is already very small.

\begin{figure}
\includegraphics[width=0.9\columnwidth]{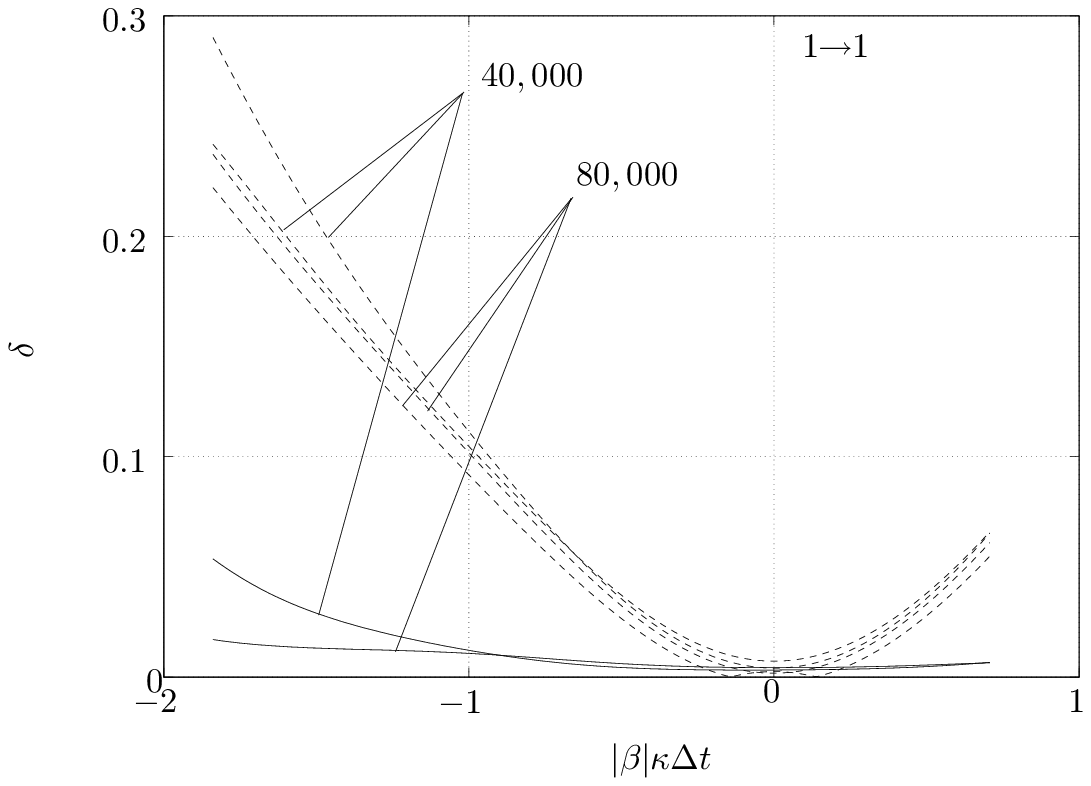}

\vspace{.35cm}

\includegraphics[width=0.9\columnwidth]{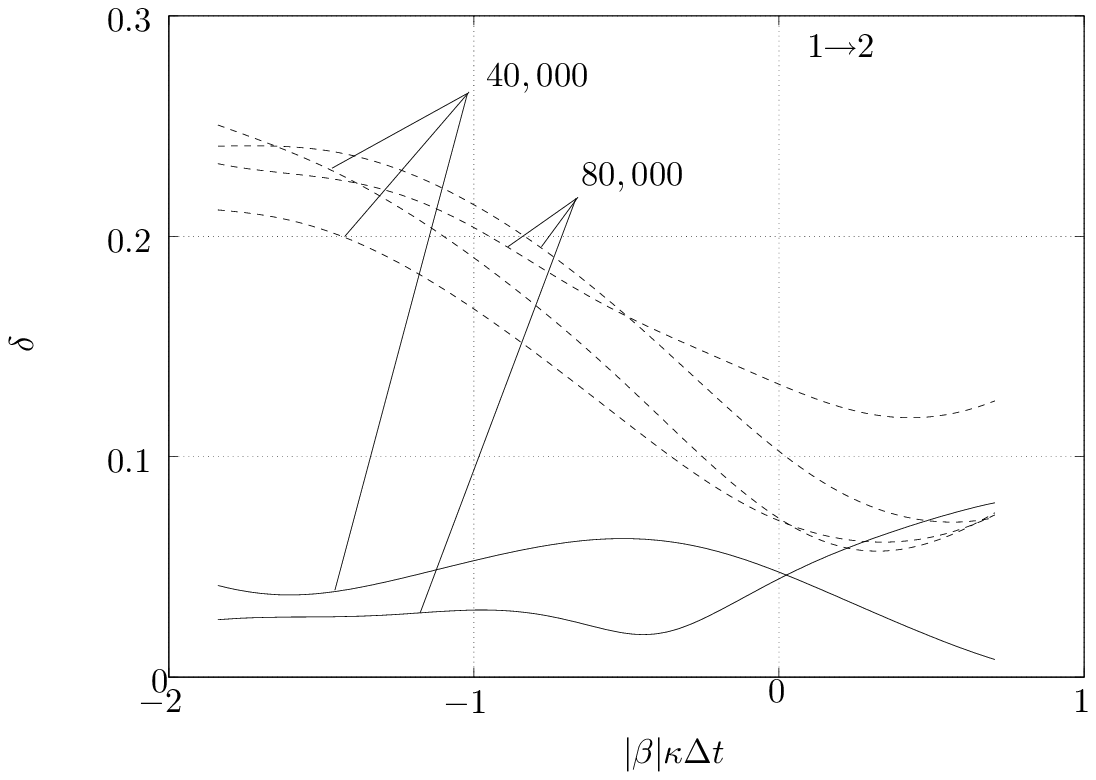}
\caption{%
Assessment of statistical errors for two mode case with $J=0.1$, $\kappa=1$, $|\beta\rangle=\sqrt{2}(1,1)$. Solid lines show the discrepancy between results derived from two independent runs with the same number of ``coin tosses'' (40,000 and 80,000). Dashed lines show the relative error of the same results compared to the exact (Hilbert-space) results. All graphs are plotted versus the scaled time $|\beta|\kappa\Delta t$, with $t_2$ set arbitrarily to $t_2=1.3$. We see that the statistical errors are insignificant.}
\label{fig:error}
\end{figure}
To develop a feeling for the statistical error, in Fig.\ \ref{fig:error} we compare the discrepancy between results obtained from two independent phase-space simulations with the relative error of the same pair of results compared to the exact (Hilbert-space) result, for the two mode Bose-Hubbard chain. The initial condition is a coherent state with $\beta=\sqrt{2}$ in each mode. The hopping strength is set to $J=0.1$, and the interaction strength to $\kappa=1$. The relative error between two correlation functions simulated in phase-space with the same number of ``coin tosses'' is shown as solid lines. The dashed lines are used for errors of the same pair of simulations in phase-space compared to the Hilbert-space solutions. The picture on top shows the said errors for the on-site average with $k=k'=1$, the one at the bottom for the averages between two modes, $k=1$, $k'=2$. All graphs are plotted versus the scaled time with $t_2$ arbitrarily chosen as $t_2=1.3$.

It is evident from the figure that the statistical errors are insignificant. The discrepancy between results in phase-space are either just small, or small compared to the accuracy of the method. At times $|\beta|\kappa\Delta t$ where the relative errors of the phase-space result compared to the Hilbert-space solution are of the same order as the errors of two independent runs in phase-space, the correlation function is already very small.

All in all, the ``response correction'' brings the results of the truncated-Wigner simulation in phase-space into a good agreement with the Hilbert-space simulations. Our observation for a single mode, that is, that the method gives good results over collapse time scales but fails on revival times, also holds for two and three modes. It is worthy of reminding the reader that all errors are solely due to the approximate nature of the truncated Wigner simulation. By itself, Eq.\ ( \ref{eq:wichtigk}) is exact.

\section{Summary}
A phase-space path-integral approach generalising the symmetric representation of {Schr\"odinger}\ operators to the Heisenberg picture is developed, and ``generalised phase-space correspondences'' allowing one to commute Heisenberg operators with unequal time arguments are derived. The conventional truncated Wigner representation emerges as an approximation within the path-integral approach. This results in formal techniques allowing one to calculate time-normal averages of {Heisenberg}\ operators approximately with relative ease. These techniques have been verified for the Kerr oscillator and for the Bose-Hubbard model showing a good agreement with exact Hilbert space calculations at collapse time scales for surprisingly low numbers of oscillator quanta. 

\section{Acknowledgements}
M.O.\ thanks the 
Institut f\"ur Quantenphysik at the Universit\"at Ulm for generous hospitality. 
L.P.\ is grateful to ARC Centre of Excellence for Quantum-Atom Optics at the University of Queensland for hospitality and for meeting the cost of his visit to Brisbane. 
The authors are indebted to D. Muth, Univ. of Kaiserslautern, for the TEBD simulations.
This work was supported by the 
Program Atomoptik of the Landesstiftung Baden-W\"urttemberg and 
SFB/TR 21 ``Control of Quantum Correlations in Tailored Matter'' 
funded 
by the Deutsche Forschungsgemeinschaft (DFG), 
a scholarship ``Mathematical Analysis
of Evolution, Information and Complexity'' at Ulm University, 
Australian Research Council, and a University of Queensland New Staff Grant. 
A.P.\ acknowledges support from AFOSR YIP and Sloan Foundation. 



\begin{thebibliography}{99}
\bibitem{EndSB}
Square brackets as a notation for the symmetric operator mapping (\ref{eq:SR}) are always signalled by a phase-space argument immediately following the brackets. In the absence of such argument square brackets have their usual mathematical meaning.
\end{thebibliography}
\end{document}